\documentclass[amsmath, amssymb,letterpaper,twocolumn,accepted=2023-03-02]{quantumarticle}
\pdfoutput=1

\usepackage{graphicx}
\usepackage{dcolumn}
\usepackage{bm}
\usepackage{dsfont} 

\usepackage{subcaption}
\usepackage{ragged2e}
\DeclareCaptionJustification{justified}{\justifying}
\usepackage[justification=justified,format=plain]{caption} 

\usepackage{graphicx}
\usepackage[colorlinks=true,citecolor=blue]{hyperref}
\usepackage[margin=1in]{geometry}

\usepackage{multirow}
\usepackage{physics}

\usepackage[capitalise]{cleveref}
\crefname{section}{Sec.}{Secs.}
\crefname{appendix}{App.}{Apps.}

\usepackage{xcolor}
\usepackage[normalem]{ulem}

\graphicspath{{figures/}}

\newcommand{\mE}{\mathcal{E}}

\newcommand{\mJ}{\mathcal{J}}

\newcommand{\mD}[2]{\mathcal{D}\left[{#1}\right] {#2} }

\newcommand{\bigoh}[1]{\mathcal{O}\left( {#1} \right)}

\newcommand{\rhoop}{ {\hat \rho}}

\newcommand{\tauutilde}{ {\hat{\tau}}^{u}}
\newcommand{\tauvtilde}{ {\hat{\tau}}^{v}}
\newcommand{\tauuvtilde}{ {\hat{\tau}}^{u/v}}

\newcommand{\tauc}{ {\tau_{c}}}

\newcommand{\sx}{\hat\sigma^{x}}
\newcommand{\sy}{\hat\sigma^{y}}
\newcommand{\sz}{\hat\sigma^{z}}

\newcommand{\ave}[1]{\langle #1 \rangle}
\newcommand{\iden}{1 \hspace{-1.0mm}  {\bf l}}

\newcommand{\Gammat}{{\widetilde{\Gamma}}}
\newcommand{\Gammap}{{\Gamma_{+}}}
\newcommand{\Gammam}{{\Gamma_{-}}}
\newcommand{\Gammapm}{{\Gamma_{\pm}}}
\newcommand{\Gammax}{{\Gamma_{x}}}
\newcommand{\Gammaperp}{{\Gamma_{x}}}

\newcommand{\Hren}{\Hop_{\rm ren}}

\newcommand{\Qop}{\hat{Q}}

\newcommand{\Aop}{\hat{A}}

\newcommand{\Aopave}{\hat{A}_{\rm avg}}

\newcommand{\Hrf}{\Hop_{\rm rf}}

\newcommand{\thetatilde}{\widetilde{\theta}}

\newcommand{\Pop}{\hat{P}}

\newcommand{\Hop}{{\hat H }}

\newcommand{\Lsop}{{\mathcal{L}}}

\newcommand{\Tsop}{{\mathcal{T}}}
\newcommand{\Vsop}{{\mathcal{V}}}
\newcommand{\Rsop}{{\mathcal{R}}}
\newcommand{\RsopTwo}{{\mathcal{R}}^{(2)}}
\newcommand{\RsopFour}{{\mathcal{R}}^{(4)}}

\newcommand{\dnorm}[1]{\norm{#1}_{\diamond}}

\newcommand{\Ci}{{\text{Ci}}}
\newcommand{\Si}{{\text{Si}}}

\begin{document}

\title{Simple master equations for describing driven systems subject to classical non-Markovian noise}

\author{Peter Groszkowski}
\affiliation{Pritzker School of Molecular Engineering, University of Chicago, Chicago, IL, USA}
\affiliation{National Center for Computational Sciences, Oak Ridge National Laboratory, TN 37831, USA}
\email{piotrekg@gmail.com}
\author{Alireza Seif}
\affiliation{Pritzker School of Molecular Engineering, University of Chicago, Chicago, IL, USA}
\author{Jens Koch}
\affiliation{Department of Physics and Astronomy, Northwestern University, Evanston, IL 60208, USA}
\author{A. A. Clerk}
\affiliation{Pritzker School of Molecular Engineering, University of Chicago, Chicago, IL, USA}

\date{April 2, 2023}

\begin{abstract}
Driven quantum systems subject to non-Markovian noise are typically difficult to model even if the noise is classical.  We present a systematic method based on generalized cumulant expansions for deriving a time-local master equation for such systems.  This master equation has an intuitive form that directly parallels a standard Lindblad equation, but contains several surprising features:  the combination of driving and non-Markovianity results in effective time-dependent dephasing rates that can be negative, and the noise can generate Hamiltonian renormalizations even though it is classical.  We analyze in detail the highly relevant case of a Rabi-driven qubit subject to various kinds of non-Markovian noise including $1/f$ fluctuations, finding an excellent agreement between our master equation and numerically-exact simulations over relevant timescales.  The approach outlined here is more accurate than commonly employed phenomenological master equations which ignore the interplay between driving and noise. 
\end{abstract}

\maketitle


\section{Introduction}
\label{sec:introduction}

As quantum technology advances, it is becoming increasingly more important to have tools for modelling dissipation and noise in driven quantum systems that are accurate, physically transparent, and easily implemented numerically.  The standard approach is to describe dissipation using a time-local Lindblad master equation (LME) \cite{Gorini_1976,Lindblad1976}; for simple systems involving a few qubits, such equations can be numerically simulated using off-the-shelf packages (see e.g.,~\cite{johansson2012qutip}).  Unfortunately, Lindblad master equations are inaccurate descriptions in many relevant situations, even if the system is just a single qubit.  If the noise is non-Markovian, and if the qubit is driven, then using a Lindblad approach generally requires an uncontrolled approximation.  

\begin{figure}[t]
    \centering
    \includegraphics[width=\columnwidth]{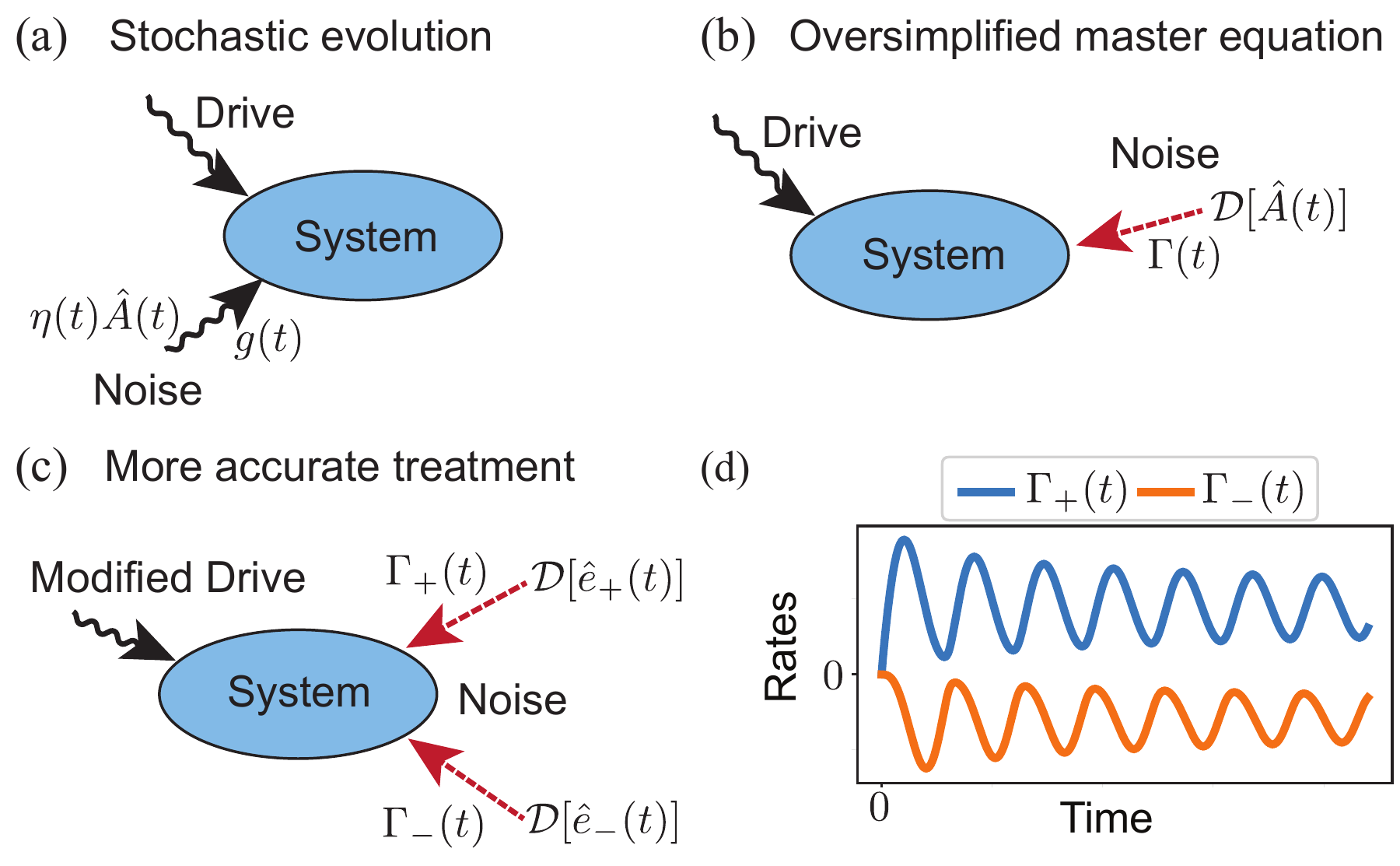}
    \caption{
        (a) Schematic of a driven qubit subject to classical, non-Markovian noise $\eta(t)$ via noise-coupling operator $\hat{A}(t)$ with strength $g(t)$. 
        (b) Standard approach to treating the noise-averaged dynamics:  one uses a master equation with a 
        single Lindblad dissipator $\mathcal{D}[\hat{A}(t)]$ 
        with corresponding rate $\Gamma(t)$ (derived in the absence of driving).  The neglect of the interplay between noise and driving can lead to large errors.     
        (c) We derive a substantially more accurate method for modeling dynamics via a pseudo-Lindblad master equation. In this case, the rates $\Gammapm(t)$, the corresponding jump operators $\hat{e}_{\pm}(t)$, and the system Hamiltonian are all modified by the drive. 
(d) Our method results in instantaneous dephasing rates that can be negative [see Eq.~\eqref{eq:lindbladMER2}]. 
   }  
    \label{fig:system}
\end{figure}

While more complicated formalisms exist for non-Markovian, driven problems (see e.g.,~\cite{breuer2002theory}), they typically lack the transparency and simplicity of an LME.  One approach is to attempt to model the effects of highly-correlated noise using an ad-hoc Lindblad master equation with either constant rates, or with rates that are explicitly time-dependent (even when the noise is stationary). These approaches are common in superconducting quantum information (see e.g.~\cite{Didier2019acFluxSweet,Blais2021Efficient,yurtalan2021characterization,Setiawan_Groszkowski_Ribeiro_Clerk_2021}), where the dominant dephasing noise typically has a $1/f$ character \cite{Ithier05,Blais_Grimsmo_Girvin_Wallraff_2020,paladino20141,paladino2002decoherence}. While these equations are often physically well-motivated and retain the simplicity of an LME, they rely on an uncontrolled approximation, so their ultimate validity is unclear. 
Other alternative attempts of capturing the effects of non-Markovian noise been also proposed (e.g., \cite{becker2021lindbladian}), however, they have not yet been widely adopted.

In this paper, we discuss a systematic, controlled approach for describing the dynamics of driven quantum systems subject to non-Markovian noise with a time-dependent but time-local master equation that is reminiscent of a LME. We focus on the general case where the dissipation is due to {\it classical} non-Markovian noise, but is still non-trivial, as the system Hamiltonian (which includes driving) does not commute with the noise Hamiltonian.  We use the machinery of cumulant expansions (well-known in other contexts, see e.g.,~\cite{kubo1962generalized,van1974cumulant,van1974cumulant2,Fox_1976}) to derive Lindblad-style master equations in this general setting.  Our approach transparently shows how the combination of non-Markovianity and driving leads to effects that are missed in more naive approaches:  the noise gives rise to a renormalization of the system Hamiltonian (even though there is no Lamb shift physics as the noise is classical), and is also described by two time-dependent rates, one of which is generically negative. As we show, this parameterization of the master equation in terms of generalized time-dependent rates provides a wealth of physical intuition.  We call this resulting equation a pseudo-Lindblad master equation (PLME)\footnote{A simialr approach has been recently also presented in \cite{gulacsi2023smoking}.}.

To illustrate and test our approach, we present a detailed study of a Rabi-driven qubit subject to various kinds of classical non-Markovian noise.  We find that even in cases where the noise has no well-defined correlation time (i.e., $1/f$ or quasistatic noise), the PLME provides an accurate description of the full dynamics over times that can be much longer than the Rabi period.  We also show that the negative rates we obtain in the PLME are in excellent correspondence to the negative rates that exactly characterize the instantaneous generator (as obtained from numerical simulation).

While cumulant expansion techniques have been employed to model classical noise in specific chemical systems~\cite{Budimir_Skinner_1987,Aihara_Sevian_Skinner_1990} and in specific qubit systems~\cite{Yang_Coppersmith_Friesen_2019}, our work differs in several respects.
In particular, the use of these techniques to give a microscopic derivation of a pseudo-Lindblad structure (with the emergence of negative dissipation rates due to colored noise and driving) is new to our work.  We note that Ref.~\cite{hall2014canonical} showed how a very general class of quantum maps can be expressed in pseudo-Lindblad form; however, unlike our work, the connection to explicit microscopic models of noise and driving was not discussed.

The remainder of this paper is organized as follows. In \cref{sec:noisyQubit}, we introduce the generic classically-stochastic quantum system of interest. In \cref{sec:cumulantGeneral}, we derive the pseudo-Lindblad master equation.  We specialize to the specific example of a Rabi-driven qubit subject to non-Markovian dephasing noise in \cref{sec:drivenQubit}. We then explore different kinds of non-Markovian noise in subsequent sections: quasistatic noise in \cref{sec:quasistatic}, noise with a Lorentzian spectrum in \cref{sec:LorentzianNoise}, and $1/f$ noise in \cref{sec:1overfNoise}. We conclude in \cref{sec:Conclusions}.


\section{General setup: Driven quantum system subject to classical non-Markovian noise}
\label{sec:noisyQubit}

Consider a generic setup where a driven finite-dimensional quantum system  (with a time-dependent system Hamiltonian) is coupled to classical noise $\eta(t)$ via a Hermitian system operator $g(t) \hat{A}$. Here, $g(t)$ represents a possibly time-dependent coupling constant. Working in the interaction picture with respect to the noise-free Hamiltonian $\hat H_0(t)$, the dynamics is described by the stochastic Hamiltonian 
\begin{align}
    \Hop(t) &=  \eta(t) g(t)\Aop(t),
    \label{eq:htotal}
\end{align}
where $\hat{A}(t) = \hat U_0(t,t_0)^\dagger \hat A \hat U_0(t,t_0)$. Here, we allow for a general time-dependent Hamiltonian $H_0(t)$ that need not commute with itself at different times. Therefore, $\hat U_0(t,t_0)=\mathcal{T}\exp(-i\int_{t_0}^t \hat{H}_0(s) ds)$, where $\mathcal{T}$ indicates the time-ordering operation and $t_0$ is an arbitrary reference time. This yields the evolution equation (setting $\hbar =1$)
\begin{align}
    \partial_{t} \rhoop_{\eta}(t) &= -i \comm{ \hat{H}(t)}{ \rhoop_{\eta}(t)},
    \label{eq:liouvEq}
\end{align}
where $\rhoop_{\eta}(t)$ is the system density matrix for a particular realization of the noise $\eta(t)$.  

In what follows, we will focus on the case where $\eta(t)$ is zero-mean Gaussian noise.  It is fully characterized by its autocorrelation function
\begin{align}
    \ave{\eta(t) \eta (t')} &=  S(t-t').
    \label{eq:Sdef}
\end{align}
Our goal is to obtain a closed, {\it time-local} evolution equation for $\rhoop(t)$, the system density matrix averaged over all realizations of the noise. This is easily done in either the case where (a) the noise is white-noise (see e.g.,~\cite{breuer2002theory}), or (b) $[\hat{A}(t),\hat{A}(t')]$ commutes with itself at all times (see~\cite{reina2002decoherence,palma1996quantum} and discussion below). These correspond respectively to the Markovian noise limit, and to the limit of undriven, pure dephasing dynamics respectively. In all other cases, deriving an accurate \textit{time-local} equation of motion for our system is a non-trivial task (despite only having classical Gaussian noise).

\section{Cumulant expansion and the pseudo-Lindblad master equation}
\label{sec:cumulantGeneral}

To systematically derive an approximate time-local equation for the noise-averaged density matrix $\rhoop(t)$, we employ the generalized cumulant expansion first introduced by Kubo \cite{kubo1962generalized}, and further developed by many others since (e.g., \cite{van1974cumulant,van1974cumulant2,Fox_1976}).
It is worth stressing that such technique is closely related to work on time-convolutionless (TCL) master equations; in particular one can think of the cumulant-expansion method as a means to systematically obtain an $n$-th order contribution to the TCL equation generator \cite{breuer2002theory,breuer1999stochastic,chaturvedi1979time}.
Thus, our approach below, can be viewed as yielding the classical-noise limit of a TCL equation (truncated at a particular order in noise strength, and rewritten in a pseudo-Lindblad form).
The technique can be viewed as yielding the classical-noise limit of the time-convolutionless master equation \cite{breuer2002theory}.
Hence, while the basic formalism presented here is discussed in many places, our main goal is to make the general structure clear, in particular the interpretation of the final equation as a pseudo-Lindblad master equation with a modified system Hamiltonian and two dephasing rates (see \cref{fig:system}). One of these rates is negative, which is a signature of the non-Markovianity of the noise process~\cite{hall2014canonical,breuer1999stochastic,breuer2002theory}.  
We point out that one common \textit{physical} interpretation for these negative rates is related to the fact that (even a classical, as is the case here) non-Markovian environment can lead to a \textit{revival} of coherence: as the system evolves, the information loss to the bath can be at times (often partially) undone and ``flow back'' into the system. In setups where the bath is quantum, this can be often heuristically understood by considering a simple model where the bath is thought of as an auxiliary mode that couples to its own (often Markovian) environment, and which also coherently couples to the original system in question. The revival of coherence can then be seen as resulting from the (partially) coherent system-auxiliary mode interaction, i.e., some of the information has a chance to traverse back the original system, before it gets permanently lost to the auxiliary mode's environment.

As is standard, we also assume that at $t=0$, the initial system density matrix $\hat{\rho}(0)$ is uncorrelated with the noise $\eta(t)$ (i.e., it is not stochastic); this singles out $t=0$, effectively breaking time-translation invariance.  

\subsection{$2^{\rm nd}$-order pseudo-Lindblad master equation}
\label{sec:PLMESecondOrder}

We start by truncating the generalized cumulant expansion at $2^{\rm nd}$-order in the noise.  We then obtain a time-local evolution equation for $\rhoop(t)$ (the noise-averaged density matrix) that has a familiar double-commutator structure.    
We can write this equation in a suggestive manner
\footnote{
This interpretation is naturally only sensible if the denominator in \cref{eq:Aopave} is non-zero for all times of interest.  For more general cases, one can simply set $\Lambda(t) = 1$, and define $\Aopave(t)$ with the denominator in \cref{eq:Aopave} set to $1$.  All of our results below then continue to apply.
}:
\begin{align}
    \partial_{t} \rhoop(t) &=   -  \Lambda(t) \comm{\Aop(t)}{ \comm{\Aopave(t)}{\rhoop(t)}},
    \label{eq:meGenR2}
\end{align}
where 
\begin{align}
    \Aopave(t) 
    &=  \frac{ 
        \int_{0}^{t} dt_{1}  S(t-t_{1}) g(t_{1}) \Aop(t_{1})  }
                {\int_{0}^{t} dt_{1}  S(t-t_{1}) g(t_{1})   }
    \label{eq:Aopave}
\end{align}
\begin{align}
    \Lambda(t) &= g(t) \int_{0}^{t} dt_{1} S(t-t_{1}) g(t_{1}),
\end{align}
Both of these quantities have a simple physical interpretation:
\begin{itemize}
    \item $\Aopave(t)$ corresponds to the weighted average of $\hat{A}(t)$ at times $t' < t$.  The weighting here is controlled by the noise autocorrelation function.    
    \item $\Lambda(t)$ is the 
        {\it instantaneous} dephasing rate we would obtain {\it if} $\hat{A}(t) = \hat{A}$ 
        (i.e.~if $\hat{A}$ commutes with $\hat{H}_0(t)$).  In that case,  $\Aopave = \hat{A}$, and \cref{eq:meGenR2} is a generalized dephasing master equation that causes coherences between eigenstates of $\hat{A}$ to decay.  For this special case, the cumulant expansion truncates at $2^{\rm}$-order and the equation is exact.
\end{itemize}

To obtain further intuition, we will now rewrite \cref{eq:meGenR2} in a form that resembles a standard Lindblad master equation, i.e., express it in terms of an effective Hamiltonian and a set of jump operators.  Given that \cref{eq:meGenR2} is trace and Hermiticity preserving, this is always possible \cite{Gorini_1976}, although the rates associated with dissipative processes will in general be time-dependent and possibly negative~\cite{breuer2002theory,hall2014canonical}.  Our goal is to show how the parameters in this pseudo-Lindblad form reflect the properties of our system and provide physical intuition.  
The first step is to quantify both the size of the operator $\Aopave(t)$, and the degree to which it is orthogonal to $\hat{A}(t)$ (the instantaneous coupling operator).  We achieve both using the Hilbert-Schmidt inner product, i.e., $\langle \hat X , \hat Y \rangle = \Tr(X^\dagger Y)$ and its corresponding norm $||\hat X||^2_{\rm{HS}} = \langle \hat X , \hat X \rangle$.  We first introduce the scalar quantities
\begin{align}
    Z_0(t) & = ||\hat A (t)||_{\rm{HS}}, \\
    Z(t) & = ||\Aopave(t)||_{\rm{HS}}, \\
    Z_\perp(t) & = 
    || 
    \Aopave(t) - \langle \hat b (t) , \Aopave(t) \rangle 
    \hat{b}(t)
     ||_{\rm{HS}},
\end{align}
as well as the normalized operators
\begin{align}
    \hat{b}(t) & = \frac{1}{Z_0(t)} \hat{A}(t), \\
    \hat{b}_{\perp}(t) & = \frac{1}{Z_\perp(t)}
    \left( 
    \Aopave(t) - \langle \hat b (t) , \Aopave(t) \rangle 
    \hat{b}(t)
\right).
\end{align}
Defining the angle $\phi(t)$ via 
\begin{equation}
    Z_\perp(t) = Z(t) \sin \phi(t),
    \label{eq:phiDefinition}
\end{equation}
we can now express $\Aopave$ in terms of operators parallel and perpendicular to the instantaneous system operator $\hat{A}(t)$:
\begin{align}
    \Aopave(t) = Z(t) \left[
        \cos \phi(t) \hat{b}(t) +
        \sin \phi(t) \hat{b}_{\perp} (t)
    \right] .
\end{align}
Thus even though our original stochastic Hamiltonian in \cref{eq:htotal} involved only a single system operator $\hat{A}(t)$ at each instant, our dynamical equation involves two independent operators:  $\hat{A}(t)$ and the orthogonal operator $\hat{b}_\perp(t)$.

Going forward, we will see that there are two time-dependent parameters controlling our master equation: the angle $\phi(t)$ in \cref{eq:phiDefinition} (which parameterizes how different $A(t)$ was at earlier times compared to the present time), and an effective rate $\tilde{\Gamma}(t)$:
\begin{align}
    \Gammat(t) &=  \frac{1}{2} \Lambda(t) Z(t) Z_0(t)
    \label{eq:Gammat}
\end{align}
This is just a normalized version of the instantaneous dephasing rate $\Lambda(t)$ introduced earlier.    

We are now in a position to express the master equation in \cref{eq:meGenR2} in a form that is formally analogous to a standard Lindblad equation.
We let $\mathcal{D}[\hat X]\hat \rho = \hat X \hat \rho \hat X^\dagger -\frac{1}{2}\{\hat X^\dagger \hat X , \hat \rho \}$ denote the standard Lindblad dissipative superoperator.   A few lines of algebra transform \cref{eq:meGenR2} into: 
\begin{align}
    \partial_{t}  \rhoop(t) =& 
   -i \comm{  \Hren(t)}{\rhoop(t)}  \nonumber \\
   & +2 \sum_{s=+,-} \Gamma_s(t) \mD{ \hat{e}_s(t) }{\rhoop(t)}  .
 \label{eq:lindbladMER2}
\end{align}
Equation~(\ref{eq:lindbladMER2}) is our general pseudo-Lindblad master equation (PLME), and is a central result of our work.  
The first term describes coherent dynamics generated by $\Hren(t)$,  a noise-induced correction to the system Hamiltonian.  It is given by the Hermitian operator:
\begin{align}
    \Hren(t) &=  -i \Gammat(t) \sin\phi(t)  \comm{\hat{b}(t)}{\hat{b}_\perp(t)}.
    \label{eq:Hren}
\end{align}
The second two terms of \cref{eq:lindbladMER2} describe generalized time-dependent dephasing processes, with Hermitian time-dependent operators $\hat{e}_\pm(t)$ that are rotated versions of $\hat{b}(t)$, $\hat{b}_\perp(t)$:
\begin{align}
    \left(\begin{array}{c}
            \hat{e}_{+}(t) \\
            \hat{e}_{-}(t) \\
    \end{array}\right) 
            &=   \left(\begin{array}{cc}
                    \cos \frac{\phi(t)}{2}  & \sin\frac{\phi(t)}{2} \\
        -\sin \frac{\phi(t)}{2} &  \cos\frac{\phi(t)}{2} \\
    \end{array}\right) 
    \left(\begin{array}{c}
            \hat{b}(t) \\
            \hat{b}_\perp(t) \\
    \end{array}\right).
    \label{eq:uvTransf}
\end{align}
The corresponding rates are given by
\begin{align} 
    \Gamma_{\pm}(t) =  \Gammat(t) \left(\cos \phi(t) \pm 1 \right).
    \label{eq:Gammapm}
\end{align}

Several comments are now in order:
The Hamiltonian $\Hren(t)$ is not a standard Lamb-shift renormalization \cite{breuer2002theory}, as there is no quantum bath here (only classical noise).  Heuristically, $2^{\rm nd}$-order perturbation theory in the stochastic Hamiltonian $\hat{H}(t)$ could yield new processes quadratic in $\eta(t)$ that have a non-zero mean. Similar terms have been discussed in specific contexts \cite{Dobrovitski2009decayofRabi,Budimir_Skinner_1987}.  Our derivation provides a more general description of such Hamiltonian renormalization terms, and shows that they are present whenever $\phi(t) \neq 0$.  

Furthermore, for any non-zero $\phi(t)$, we always have two effective dephasing processes involving two orthogonal operators $\hat{e}_\pm(t)$. This reflects the fact that dissipation involves both the system operator $\hat{A}(t)$ at the current time, as well as at earlier times.  For small $\phi(t)$, these operators are almost the same as $\hat{b}(t)$ and $\hat{b}_\perp(t)$. 

Finally, for any non-zero $\phi(t)$ the effective instantaneous dephasing rate $\Gamma_{-}(t)$ is always negative.
While surprising, such negative dephasing rates are known to arise in problems where dephasing dynamics is non-monotonic in time \cite{Breuer2016RMP}.  As we will show via comparison to full numerical simulations, the negative dephasing rate we predict is not an artefact of our approximation, but is instead in excellent agreement with features of the exact dissipative evolution.

\subsection{Simple limits}

It is reassuring to confirm that \cref{eq:lindbladMER2} reduces to known results in some simple limits.  First, consider the limiting case of Markovian noise, where the noise autocorrelation function approaches a $\delta$-function: $S(t) \rightarrow \sigma \delta(t)$.  In this case, the bath has no memory, and therefore $\Aopave(t) = \hat{A}(t)$.  As a result, $\Aopave(t)$ has no component orthogonal to $\hat{A}(t)$, hence $Z_\perp(t)$ and $\phi(t)$ vanish.   
Our pseudo-Lindblad master equation thus reduces to:
\begin{align}
   \partial_{t} \rhoop(t) &= \sigma
   \mD{ g(t) \hat{A}(t) }{\rhoop(t)}.
 \label{eq:lindbladMEwhite}
\end{align}
This is a Lindblad master equation describing generalized dephasing with an instantaneous rate $\sigma g(t)^2$.  In this white-noise limit, this equation is exact, as there are no higher-order cumulants (as the lack of any memory means the dynamics is insensitive to whether $\hat{A}(t)$ commutes with itself at different times).

A second simple case is that of a time-independent $\hat{A}(t)$.  Again, in this case $\phi(t) = 0$,
and we recover the expected dephasing Lindblad equation:
\begin{align}
        \partial_{t} \rhoop(t) &= 
        4 \tilde{\Gamma}(t) \mathcal{D}[\hat{b}] \rhoop(t)
        =   2 \Lambda(t) \mathcal{D}[\hat{A}] \rhoop(t),
\end{align}
which is also exact in this case, as all higher cumulants vanish.

\section{Example of the formalism: Rabi driven qubit subject to non-Markovian noise}
\label{sec:drivenQubit}

\subsection{Form of the PLME}

Next, we apply the general results derived above to an ubiquitous but surprisingly rich problem:  a Rabi-driven qubit that is subject to classical non-Markovian dephasing noise. Treating the driving in the rotating-wave approximation, the rotating frame Hamiltonian of the system is 
\begin{align}
     \Hrf(t) &= \frac{\Omega(t)}{2} \sx + \eta(t)\sz
     = \hat{H}_0(t) +  \eta(t)\sz,
     \label{eq:hrf}
\end{align}
where $\Omega(t)$ is the Rabi drive amplitude. This system the form of our general problem, with $g(t)= 1$. 

The corresponding interaction-frame system operator $\Aop(t)$ is then
\begin{align}
    \Aop(t) &=  \cos(\theta(t))\sz + \sin (\theta (t)) \sy ,
    \label{eq:AopDrivenQubit}
\end{align}
with 
\begin{align}
    \theta(t) &=  \int_{0}^{t}dt' \Omega(t').
    \label{eq:theta}
\end{align}
Note that our treatment below goes beyond previous work on this problem that focused exclusively on the slow-noise limit \cite{Dobrovitski2009decayofRabi,rabenstein2004qubit}.

We can now directly apply the results of \cref{sec:cumulantGeneral}. 
Using the $\Aop(t)$ given by \cref{eq:AopDrivenQubit}, the PLME (based on keeping contributions up to $2^{\rm nd}$-order in the noise strength) follows from \cref{eq:lindbladMER2}.  It takes a transparent form in the rotating frame used to write $H_0(t)$ in \cref{eq:hrf}:
\begin{align}
    \partial_{t}  \rhoop(t) =& 
   -i \comm{  \Hren(t)}{\rhoop(t)}  
   +\Gammap(t) \mD{ \tauutilde(t) }{\rhoop(t)}  \nonumber \\
   &+\Gammam(t) \mD{ \tauvtilde(t) }{\rhoop(t)}. 
 \label{eq:lindbladMER2DrivenQubit}
\end{align}
The instantaneous dephasing $\Gammapm(t)$ are given by \cref{eq:Gammapm}, while from \cref{eq:Hren}, the noise-induced Hamiltonian correction is 
\begin{align}
    \Hren(t) &=  - \Gammat(t) \sin\phi(t) \sx .
    \label{eq:HrenDrivenQubit}
\end{align}
Here $\phi(t)$ is given by \cref{eq:phiDefinition}.
The corresponding jump operators $\tauuvtilde(t)$ are rotated Pauli matrices, and read
\begin{align}
    \left(\begin{array}{c}
            \tauutilde(t) \\
            \tauvtilde(t) \\
    \end{array}\right) 
            &=   \left(\begin{array}{cc}
            \cos \thetatilde(t)  & \sin\thetatilde(t) \\
        -\sin\thetatilde(t) &  \cos\thetatilde(t) \\
    \end{array}\right) 
    \left(\begin{array}{c}
            \sz \\
            \sy \\
    \end{array}\right),
    \label{eq:uvTransfDrivenQubit}
\end{align}
with
\begin{align}
    \thetatilde(t) = \frac{\phi(t)}{2} + \theta(t).
    \label{eq:thetatilde}
\end{align}
Note that $\tauuvtilde(t)$ are the same as $\sqrt{2} \hat{e}_{\pm}(t)$ in the general PLME \cref{eq:lindbladMER2}, but expressed in the rotating frame used to write \cref{eq:hrf}.

In the next sections we will explore the accuracy of the above approximate master equation, by considering its behavior given specific non-Markovian noise correlation functions. 

\subsection{Corrections to the PLME from $4^{\rm th}$-order cumulants}
\label{sec:PLMEFourthOrder}
We can further improve the accuracy of our PLME by including higher-order noise terms that arise at $4^{\rm th}$-order in the cumulant expansion. The $2^{\rm nd}$-order equation, \cref{eq:lindbladMER2DrivenQubit}, has two dephasing dissipators involving Pauli matrices in the $z-y$ plane. As one might expect, $4^{\rm th}$-order contributions modify the rates and forms of these dissipators.  There is, however, a more interesting effect that emerges at $4^{\rm th}$-order:  a new, third dissipator that corresponds to dephasing along the $x$ axis (again, working in the rotating frame used to write \cref{eq:hrf}).  This process corresponds to the term:
\begin{equation}
    \Gamma_x(t) \mathcal{D} [ \sx ] \rhoop(t).
    \label{eq:4thorderx}
\end{equation}

In \cref{app:CumulantBasics} we give the general expression for the $4^{\rm th}$-order cumulant contribution to the equation of motion, which in turn, lets us calculate the explicit form of $\Gamma_{x}(t)$.
We stress that this new term has a simple heuristic explanation. We saw that at $2^{\rm nd}$ order noise generates a renormalization of the coherent system Hamiltonian $\Hren(t) \propto \sx$, i.e., a correction to the effective Rabi drive amplitude. Not surprisingly, there are fluctuations in this noise-induced change of the drive amplitude.  The dissipator in \cref{eq:4thorderx} simply describes the dephasing associated with these fluctuations.


\section{Rabi driven qubit subject to quasistatic noise}
\label{sec:quasistatic}

In this section, we apply the PLME description to a Rabi-driven qubit subject to longitudinal classical noise (see~\cref{eq:hrf}), and study what is perhaps the most non-Markovian of all Gaussian noise processes: quasistatic noise, with an autocorrelation function that is independent of time:
\begin{align}
    S(t) = \sigma^{2}.
    \label{eq:Squasi}
\end{align}
While highly non-Markovian (i.e., noise correlations have an infinite time correlation), this noise is especially simple to work with, and results in compact analytical expressions. 
It also makes benchmarking our approach against exact numerics easy
(as the propagator for any single noise realization can be calculated exactly).

Note that in the absence of a finite correlation time (i.e., exponential decay of $S(t)$ at long times), it is well known that the generalized cumulant expansion does not converge if one considers extremely long evolution times \cite{van1974cumulant,van1974cumulant}.  This deficiency is however not an issue for most quantum information applications, as one wants to understand finite-time evolution (i.e., during the execution of a gate or computation), and not some long-time dissipative steady state.  For such short-to-intermediate timescales, the cumulant expansion can still provide an accurate description of the dynamics, whenever noise is not especially strong (something we show explicitly in what follows by comparing against numerically-exact results).

\subsection{PLME parameters}

\subsubsection{Rates and Hamiltonian renormalization}

For simplicity, we assume a constant Rabi drive amplitude, $\Omega(t)=\Omega$.   We then find that the two instantaneous dephasing rates in the $2^{\rm nd}$-order PLME of \cref{eq:lindbladMER2DrivenQubit} are given by \cref{eq:Gammapm}:
\begin{align}
    \Gammapm(t) &= \frac{\sigma^{2}}{\Omega} \left( \sin (\Omega t ) 
    \pm\ 2 | \sin (\Omega t / 2) | \right),
    \label{eq:GammapmQuasi}
\end{align}
while the angle $\phi(t)$ that controls the form of the two dephasing dissipators is simply given by
$\phi(t)=-\Omega t/2$. We see that both dephasing rates have a simple oscillatory form, with $\Gamma_+(t) \geq 0$ and $\Gamma_-(t) \leq 0$.  Further, both rates vanish periodically at times $t_n = 2 \pi n / \Omega$.  This reflects the basic physics of continuous dynamical decoupling \cite{Lidar2004Unification,Retzker2012RobustDynamical,Awschalom2020Universal}:  the Rabi drive tends to periodically average out the dominant dephasing effects of the noise.

The remaining noise-induced term in the $2^{\rm nd}$-order PLME is the Hamiltonian renormalization
\begin{align}
    \Hren(t) &= \frac{\sigma^{2}}{\Omega} \left(1 -  \cos (\Omega t) \right) \sx .
\end{align}
Similar to the dephasing effects, we see that this noise-induced Hamiltonian also vanishes periodically in time.  

We can also compute the dominant symmetry breaking term that emerges from the $4^{\rm th}$-order cumulant corrections to the PLME, i.e., the induced $x$ dephasing term in \cref{eq:4thorderx}.  We find that the rate of this process is always positive, but not periodic:
\begin{align}
    \Gammaperp(t) =& \frac{2 \sigma^{4} }{\Omega^{3}} 
    \big(2 \Omega t 
        -2 \Omega t \cos (\Omega t )  \nonumber \\
    &+ \sin (2 \Omega t) -2 \sin (\Omega t)\big).
    \label{eq:quasiGammaPerp}
\end{align}

\subsubsection{Limiting forms for weak driving and/or short times}

It is instructive to examine the form of the PLME parameters in the short-time or weak-drive limit, $\Omega t \ll 1$.  In this case, we find  (dropping terms $\bigoh{(\Omega t)^3}$)
\begin{align} 
    \Gamma_{+}(t) &\approx 2\sigma^{2}t \left(1 -\frac{5}{48} ( \Omega t)^{2}\right),
    \label{eq:quasGammapSmallA}
\end{align}
\begin{align} 
    \Gamma_{-}(t) &\approx - 2\sigma^{2}t \frac{(\Omega t)^{2}}{16},
    \label{eq:quasGammamSmallA}
\end{align}
while $\Hren(t)$, reduces to 
\begin{align} 
    \Hren(t) &\approx \frac{ \Omega \sigma^{2} t^{2}}{2}\sx . 
    \label{eq:quasHrenSmallA}
\end{align}

To $0^{\rm th}$-order in the drive amplitude, both the negative dephasing rate and the Hamiltonian correction vanish, and we are left with a single, linear-in-time positive dephasing rate.  This corresponds to the expected generalized dephasing rate for the undriven problem (a limit where, as discussed, our equation becomes exact).  
At $2^{\rm nd}$ order in the drive, things become more interesting:  the magnitude of $\Gamma_{+}(t)$ is suppressed, and we generate a non-zero negative dephasing rate $\Gamma_{-}(t)$.  

\subsection{Comparison of instantaneous dissipation rates: PLME vs numerically-exact results}
\label{sec:QuasiExactRates}

\begin{figure}[t]
    \centering
    \includegraphics[width=0.46\textwidth]{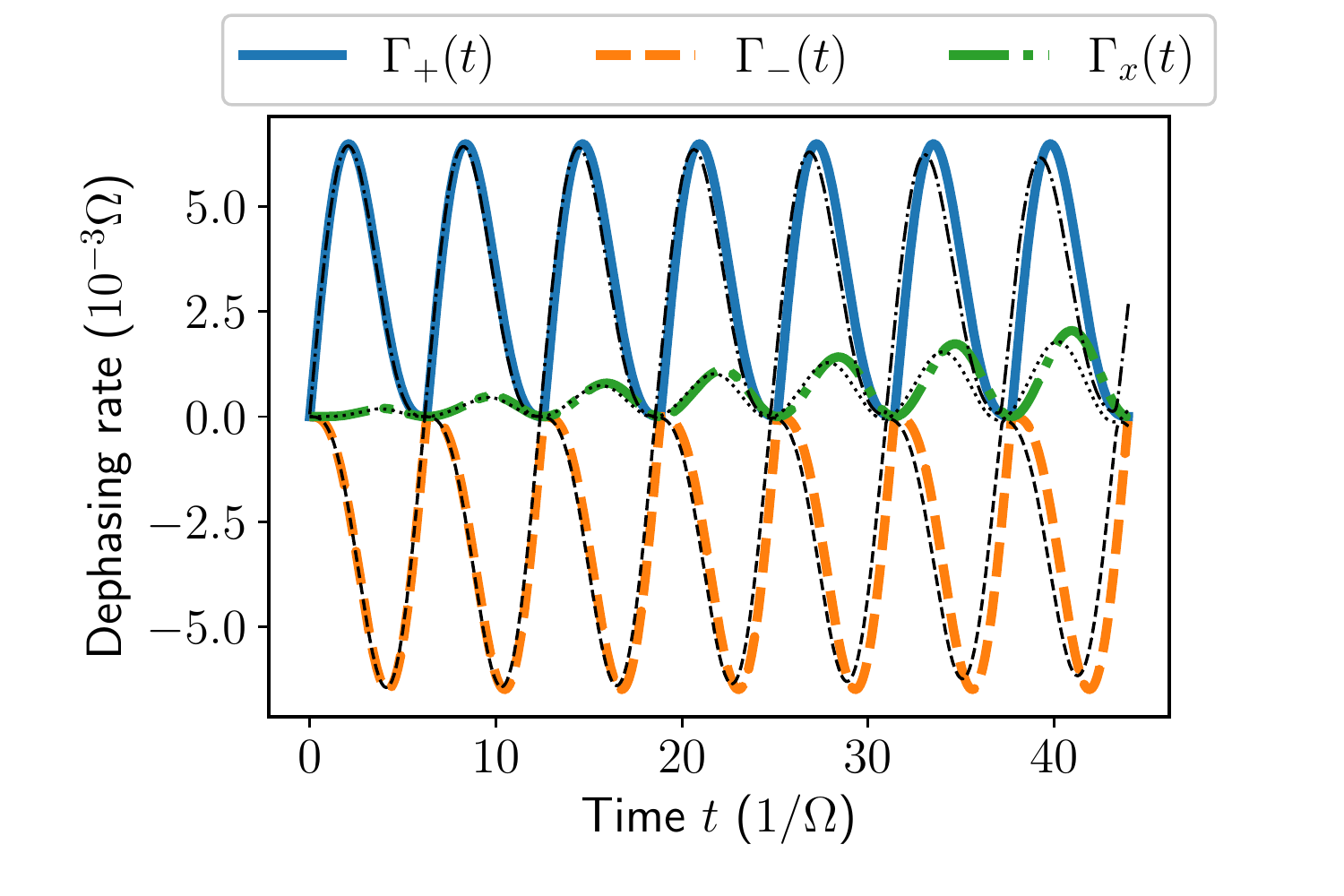}
    \caption{
        Time evolution of $\Gammapm(t)$ (blue and orange) from the $2^{\rm nd}$-order PLME of \cref{eq:lindbladMER2}, as well as the rate $\Gamma_{x}(t)$ (green), calculated by keeping the $4^{\rm th}$-order noise contributions in the cumulant expansion. 
        For comparison, the black curves are the rates obtained from the \textit{exact} system evolution, showing good agreement with our analytically derived approximations. We especially stress that that the large-in-magnitude negative rate $\Gammam(t)$ is present in the true dynamics of the system. In the simulations, we take $\sigma/\Omega=0.05$. 
    }
    \label{fig:quasiRates}
\end{figure}

While negative dephasing rates may be alarming at first, they are a common feature of non-Markovian dynamics~\cite{hall2014canonical}. To instill confidence, we numerically calculate the instantaneous rates that characterize the full dissipative evolution of our system (without any approximation).  
The definition of such rates is discussed in many places (e.g.,
Refs.~\cite{hall2014canonical,Breuer2016RMP}).
We numerically calculate the full quantum map $\Vsop(t)$ that propagates $\rhoop(0)$ to $\rhoop(t)$, and then obtain the instantaneous generator of the dynamics via \cite{Breuer2016RMP}  
\begin{align}
    \Lsop_{\rm inst}(t)  & \approx \lim_{dt \rightarrow 0}  \frac{1}{dt} \left( \Vsop (t+dt) \Vsop^{-1}(t) - \iden \right).
\end{align}
As was done in \cref{sec:cumulantGeneral}, $\Lsop_{\rm inst}(t)$ can be brought into a Lindblad-like form. This lets us extract the generalized instantaneous dephasing rates that characterize the full dynamical map.

Figure \ref{fig:quasiRates} shows the results for our analytical predictions for $\Gammapm(t)$ (orange and blue curves) which closely reproduce the two large-in-magnitude rates obtained from the exact evolution (black curves). We stress the central point of these results: the highly negative $\Gammam(t)$ that is present in our PLME, is also there in the \textit{true} evolution. As it is as large in amplitude as $\Gammap(t)$, its presence can be important in accurately modeling the qubit's evolution.
Furthermore, is worth stressing that the exact dynamics also predicts a third rate (black dotted curve), which is accurately captured by $\Gammaperp(t)$, obtained from keeping the $4^{\rm th}$-order terms in the cumulant expansion.

\subsection{Benchmarking the PLME against numerically-exact results}
\label{sec:QuasiPerformanceComparison}

In order to asses the usefulness of the PLME, we need a means to quantify its accuracy against the \textit{exact} system evolution.  For our purposes, we will compare against the {\it numerically-exact} evolution, obtained by numerically averaging unitary evolution 
over realizations of the stochastic process $\eta(t)$.  

There are various metrics that can be used to assess the accuracy of an approximate open quantum systems evolution.  One approach is to consider the performance of the approximation only for a small, specific subset of initial states \cite{Zhang_Pokharel_Levenson-Falk_Lidar_2021}.  A more complete approach is to try to compare the full structure of the exact and approximate evolution maps (e.g., Ref. \cite{Hartmann_Strunz_2020}).

Here, we will focus on a measure that characterizes the entire evolution map, and is not tied to a particular choice of initial states (although in \cref{app:expvals}, for completeness, we also show the evolution of specific Pauli operator expectation values given a particular initial state). For some evolution time $t$, let $\mJ_{\rm exact}(t)$ denote the Choi matrix of the exact evolution map, and $\mJ_{\rm approx}(t)$ the Choi matrix of the evolution map obtained via the approximation of interest (i.e.,~the PLME). We then define the \textit{approximation error} at time $t$ by:
\begin{align}
    \epsilon(t) &= \dnorm{ \mJ_{\rm exact}(t) - \mJ_{\rm approx}(t)}.
    \label{eq:errorDnorm}
\end{align}
Here 
$\dnorm{...}$ is the diamond norm~\cite{aharonov1998quantum,watrous2018theory}.  We choose the diamond norm to measure the size of the deviation between the approximate and true maps, as this gives  $\epsilon(t)$ a direct operational meaning: it is directly tied to the single-shot channel distinguishability between these two maps \cite{watrous2018theory}.  $\epsilon$ is 0 when the map associated with the \textit{approximate} dynamics cannot be distinguished from the \textit{exact}  map after a single evolution (no matter how the initial state is optimized), and equals 2 when the two maps are maximally distinguishable \cite{watrous2018theory}.  At a heuristic level, it characterizes the worst case performance of our approximation, i.e.,~the largest error maximized over all possible initial states. 

\begin{figure}[t]
    \centering
    \includegraphics[width=0.46\textwidth]{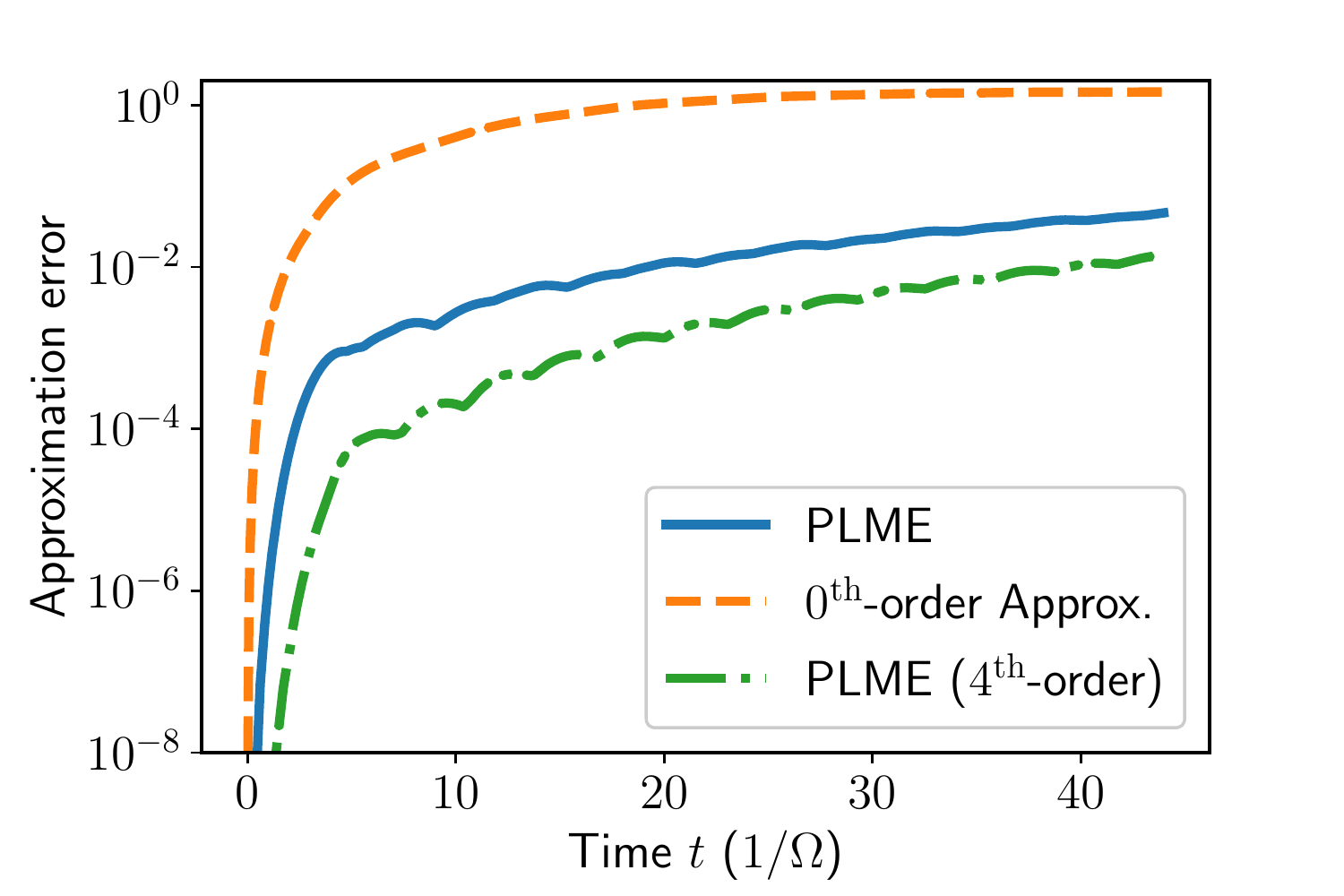} 
    \caption{
        The accuracy of different approximations for calculating the qubit's evolution under quasistatic noise. The approximation error of \cref{eq:errorDnorm} is shown as a function of the evolution time for the PLME when $2^{\rm nd}$-order (blue) or $4^{\rm th}$-order (green) terms are kept in the cumulant expansion, as well as the for the $0^{\rm th}$-order approximation of \cref{eq:quasi0th} (orange).  We see that both versions of the PLME yield large accuracy improvements over the naive $0^{\rm th}$ equation.  All curves correspond to $\sigma/\Omega=0.05$. 
    }
    \label{fig:quasiError}
\end{figure}

With this definition in hand, we can now quantify the approximation error of the PLME for different choices of system parameters and noise spectra.  To put these results in context, we will also consider the approximation error associated with a much simpler kind of time-local master equation.  In this approach (which we will term the $0^{\rm th}$-order master equation), one obtains a Lindblad-form master equation in two steps.  First, one ignores the Rabi driving, and derives a dephasing dissipator (with an $\Omega$-independent rate).  Then, one simply puts back the Rabi driving in the Hamiltonian term of the master equation.  

Working in the interaction frame (and focusing on quasistatic noise), the 
$0^{\rm th}$-order master equation takes the form:
\begin{align}
    \dot \rhoop(t) &=  2 \sigma^{2} t \mD{\cos (\Omega t) \sz + \sin(\Omega t) \sy}{\rhoop}(t),
    \label{eq:quasi0th}
\end{align}
We stress that this kind of approach is commonly used because of its simplicity \cite{Didier2019acFluxSweet,Blais2021Efficient,yurtalan2021characterization,Setiawan_Groszkowski_Ribeiro_Clerk_2021}.  Unlike the PLME, it results in only a single, $\Omega$-independent dephasing rate.

The approximation error $\epsilon(t)$ of the PLME as well as the naive $0^{\rm th}$-order master equation of \cref{eq:quasi0th} are shown in \cref{fig:quasiError}, 
for quasistatic noise that is weak compared to the Rabi frequency, $\sigma / \Omega = 0.05$. 
For comparison, besides only looking at the $2^{\rm th}$-order PLME explicitly shown in \cref{eq:lindbladMER2DrivenQubit}, here, for quasistatic noise, we also include the performance of a PLME obtained by keeping all the $4^{\rm th}$ order contributions, which result in: corrections to the rates $\Gammapm(t)$, the corresponding jump operators (that rotate in the $z-y$ plane), the Hamiltonian renormalization $\Hren(t)$, and importantly, the $x$-dephasing rate $\Gammax(t)$, and its jump operator.

We see that both forms of the PLME (obtained through either $2^{\rm nd}$- or $4^{\rm th}$-order cumulant expansions) 
have a significantly smaller approximation error than the $0^{\rm th}$-order equation. 
As expected, $\epsilon(t)$ for both schemes grows with increasing time.  Nonetheless, for weak noise, the PLME approximation error remains low even if $t \gg 1 / \Omega$.
To be concrete, we see that for the chosen parameters, the $0^{\rm th}$-order approach has an approximation error $\epsilon(t) < 10^{-3}$ only if $t \Omega \lesssim 1$. 
In contrast, the $2^{\rm th}$-order PLME has $\epsilon(t) < 10^{-3}$ at times where $t\Omega \lesssim 5$, while the $4^{\rm th}$-order PLME, over even longer longer times, where $t\Omega \lesssim 17$.
We also point out that in \cref{app:expvals}, we present an example of the expectation values evolution for a particular initial state.

\begin{figure}[t]
    \centering
    \includegraphics[width=0.46\textwidth]{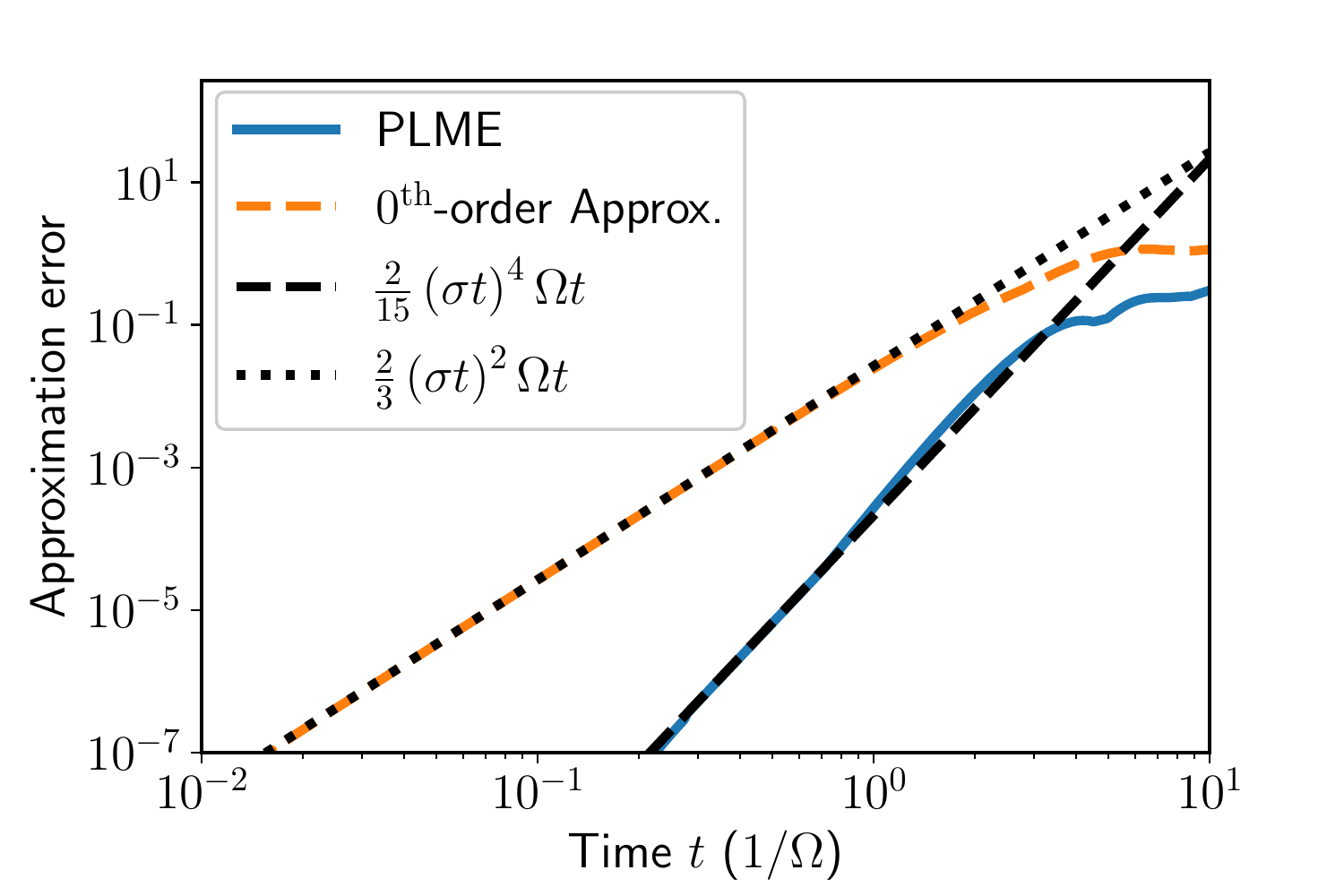}
    \caption{
        Approximation errors $\epsilon(t)$
        of the $2^{\rm nd}$-order PLME (blue) and the $0^{\rm th}$-order approximation (orange) for a Rabi driven qubit subject to quasistatic noise, in the short time limit $\Omega t \ll 1$. 
        Black dotted and dashed curves show $\epsilon(t)$ associated with the $0^{\rm th}$-order approximation scales as $t^3$.  In contrast, for the PLME is scales as $t^5$. One also finds a different asymptotic scaling with the noise strength $\sigma$ ($\sigma^2$ for the $0^{\rm th}$-order equation, $\sigma^4$ for the PLME.  These results demonstrate the gain in accuracy associated with using the PLME). All curves correspond to $\sigma/\Omega=0.2$.
    }
    \label{fig:quasiErrorShortTime}
\end{figure}

For further insights, we can also consider the behavior of $\epsilon(t)$ for the different approaches in the $t \rightarrow 0$ limit. Unfortunately, the asymptotics of the diamond norm are difficult to compute. As a proxy, we can instead consider a different error measure 
where the diamond norm is approximated by the 1-norm 
\footnote{We consider a 1-norm of a superoperator associated with the difference between the average evolution propagators of the maps representing the two processes being compared. Here, a 1-norm of a superoperator $\Vsop(t)$ is defined as: $||\Vsop(t)||_{1} = {\text{max}}_{j} \sum_{i} \abs{v_{ij}(t)}$, with $v_{ij}(t)$ representing the $ij$-th element of $\Vsop(t)$.}.
We find from numerics that this quantity behaves analogously to $\epsilon(t)$ in the $t \rightarrow 0$ limit.  Further, the 1-norm metric is amenable to analytic estimates.  
Using this equivalence, we find that for short times, the approximation error 
for the $2^{\rm nd}$-order PLME scales as:
\begin{align}
\mathbb{\epsilon}_{\rm PLME}(t) &\approx  \frac{2}{15} (\sigma t)^{4} \Omega t \label{eq:quasiShortPLME},
\end{align}
whereas for the naive $0^{\rm th}$-order equation we obtain
\begin{align}
\mathbb{\epsilon}_{0^{\rm th}}(t) &\approx \frac{2}{3} (\sigma t)^{2} \Omega t. 
\label{eq:quasiShort0th}
\end{align}
From \cref{fig:quasiErrorShortTime} we see that \cref{eq:quasiShortPLME,eq:quasiShort0th} accurately described the asymptotic behavior of $\epsilon(t)$ in the small $t$ limit.   
The more favorable approximation error scaling of the PLME is not surprising, as in contrast to the $0^{\rm th}$-order approximation of \cref{eq:quasi0th}, it fully captures $2^{\rm nd}$-order noise contributions to the qubit's dynamics.

Finally, while the PLME yields a highly accurate approximation, its form is not guaranteed to generate a completely positive map. In particular, for certain initial conditions, the evolved density matrix may acquire small, negative eigenvalues.  The size of this unphysical negativity is necessarily small for short times and weak noise, as it is bounded by the error metrics we have considered above.  We discuss the positivity of our approximation in more detail in \cref{app:Positivity}.  

\section{Rabi driven qubit subject to noise with a finite correlation time}
\label{sec:LorentzianNoise}

\begin{figure*}[t]
    \centering
    \begin{subfigure}{0.990\textwidth}
        \caption{}
    \includegraphics[width=\textwidth]{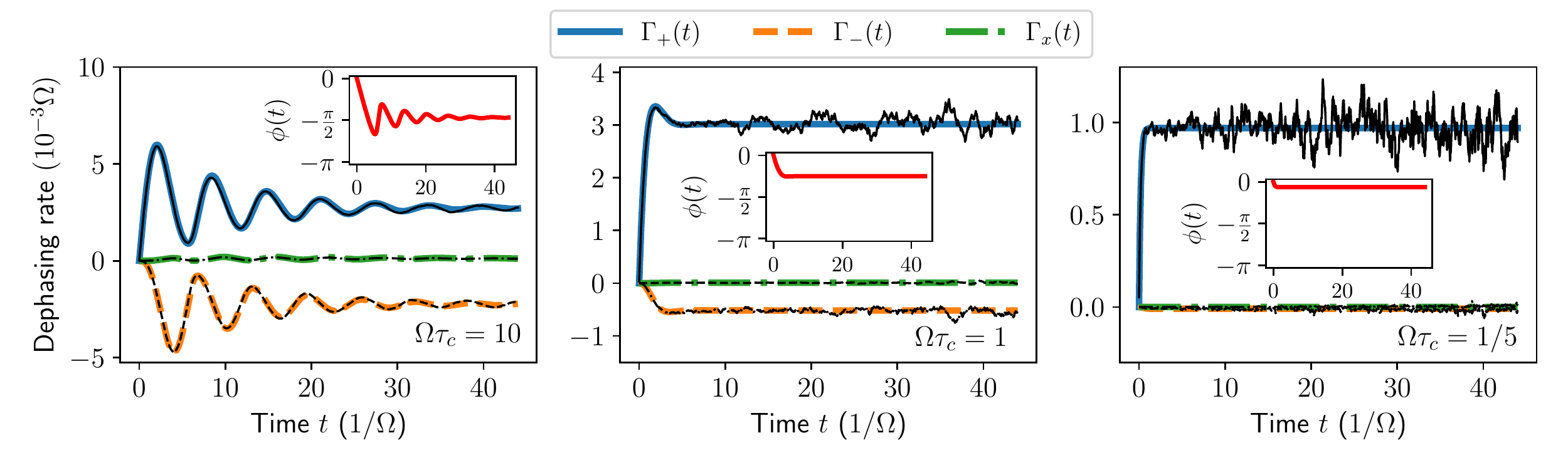}
    \end{subfigure} \\
    \begin{subfigure}{0.990\textwidth}
        \caption{}
    \includegraphics[width=\textwidth]{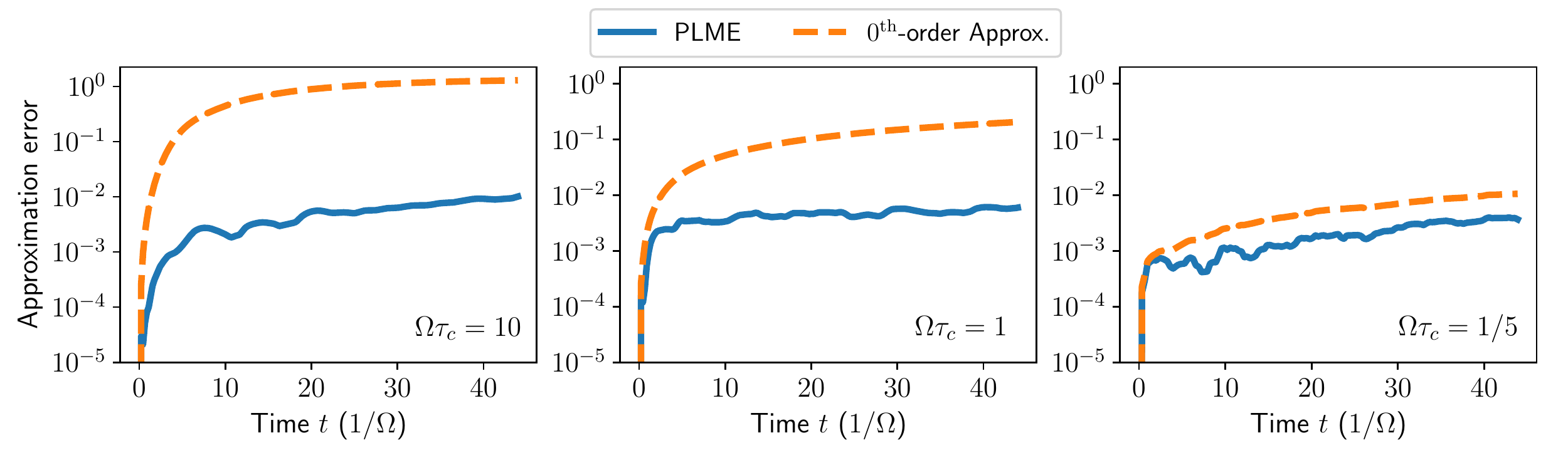} 
    \end{subfigure} 
    \caption{
        Accuracy and performance of the PLME for the case of Lorentzian noise.
        (a) $2^{\rm nd}$-order PLME rates $\Gammapm(t)$ (blue and orange) and $\Gammax(t)$ (green curve), obtained by keeping $4^{\rm th}$-order terms in the cumulant expansion (see \cref{app:Gammaxexpressions}).
        The correlation time $\tauc$ is largest in the left panel and decreases towards the right. In particular, we have $\Omega \tauc = 10$, (left column), $\Omega \tauc =1$ (middle column) and $\Omega \tauc=1/5$ (right column).
        Black curves in each plot present rates obtained from the true system evolution (here estimated by averaging over 20,000 noise realizations). 
       The time evolution of the angle $\phi(t)$ (red) is shown in the insets. 
        As the noise becomes less correlated (i.e., $\Omega\tauc$ decreases), all rates and $\phi(t)$ show fewer oscillations and quickly approach constant values. Furthermore, the negative rate $\Gammam(t)$ becomes smaller and less relevant, thus making $\Gammap(t)$ a dominant source of incoherent evolution. 
        (b) The approximation error of \cref{eq:errorDnorm} as a function of time quantifying the accuracy of the evolution obtained from the $2^{\rm nd}$-order PLME (blue) or the $0^{\rm th}$-order approximation of \cref{eq:lorentzZerothOrder} (orange). 
        In all plots the noise strength is taken to be $\sigma=0.05\Omega$. 
    }
    \label{fig:lorentzErrorRatesCombined}
\end{figure*}

In this section we study the Rabi-driven qubit (see~\cref{eq:hrf}), now subject to classical dephasing noise that has a finite correlation time $\tau_c$.  We take $\eta(t)$ to be an Ornstein-Uhlenbeck process, implying:
\begin{align}
    S(t) &=  \sigma^{2} \exp \left( - |t| / \tauc \right)
    \label{eq:LorentzS}.  
\end{align}
This yields a Lorentzian noise spectral density as encountered generically in various settings
(i.e., classical telegraph noise produced by a two-state fluctuator, in the Gaussian approximation).
It also provides a convenient test case, as we can interpolate between the quasistatic noise of the previous section, and noise that is Markovian, simply by tuning $\tau_c$.

\subsection{PLME parameters}

The $2^{\rm th}$-order PLME of \cref{eq:lindbladMER2DrivenQubit} has the effective dephasing rates given by:
\begin{align} 
    \Gammapm(t) =& 
    \frac{\sigma^{2}\tauc}{(\Omega \tauc)^{2} + 1 }
    \bigg( e^{-t/\tauc} (\Omega \tauc \sin(\Omega t) - \cos(\Omega t))          \nonumber \\
        & + 1 \pm \sqrt{(\Omega\tauc)^{2} + 1 } 
        \nonumber \\
    & \times  
(e^{-2t/\tauc} - 2 e^{-t/\tauc} \cos(\Omega t) + 1)^{\frac{1}{2}}\bigg).
\label{eq:lorentzGammas}
\end{align}
The jump operators corresponding to these rates $\tauutilde(t),\tauvtilde(t)$ are determined by \cref{eq:uvTransfDrivenQubit,eq:thetatilde}, with the non-Markovian angle $\phi(t)$ satisfying
\begin{align} 
    \cos\phi(t)&=   
\frac{ e^{-t/\tauc} (\Omega \tauc  \sin(\Omega  t )-\cos(\Omega  t ) ) + 1}
{K},
\end{align}
and
\begin{align} 
\sin\phi(t) &= \frac{ e^{-t/\tauc }   ( \Omega  \tau_{c}\cos (\Omega  t )  +\sin (\Omega  t )) -\Omega  \tau_{c} }
{K},
\end{align}
with 
\begin{align}
    K&=     \sqrt{(\Omega^{2} \tau_{c}^{2}+1) (1-2 \cos (\Omega  t ) e^{-t/\tauc}+e^{-2t/\tauc}}).
\end{align}
Note that $\phi(t)$ exhibits initial oscillatory behavior, but for times longer than the correlation time (i.e., $t \gg \tauc$), it has the simple form $\phi(t) \approx -\Omega \tauc$. Hence, for noise that is only very weakly correlated, we see that $\thetatilde(t) \approx \Omega (t - \tauc/2)$; the jump operator associated with the rate $\Gammap(t)$ is equivalent to the system operator $\Aop(t)$, but acting at a slightly ``earlier time''.

We also obtain a Hamiltonian renormalization
\begin{align} 
\Hren(t) =& \frac{ (\sigma \tauc)^{2} \Omega }{ (\Omega \tauc)^{2}+ 1}  \nonumber  \\
 & \times \left( 1 - e^{-t/\tauc} \left(  \cos (\Omega t ) - \frac{\sin (\Omega t )}{\Omega\tauc}  \right) \right) \sx .
\end{align}
In the $\tauc \rightarrow \infty$ limit, one recovers the results for quasistatic noise presented in \cref{sec:quasistatic}. In contrast, one can obtain the Markovian white-noise limit by taking $\sigma^2 = D / \tau_c$ and letting $\tau_c \rightarrow 0$.  In this limit $\Gamma_{-}(t)$ and $\Hren(t)$ vanish as expected.

Equation (\ref{eq:lorentzGammas}) implies that a $2^{\rm nd}$-order PLME, for any non-zero correlation time, we still have two instantaneous dephasing rates $\Gammapm(t)$, with $\Gamma_{-}(t) < 0$. Similarly to the case of quasistatic noise, at $4^{\rm th}$-order, one also observes a third rate $\Gammax(t)$, that corresponds to dephasing along the qubit's $x$-axis. We explicitly show the analytical form of $\Gammax(t)$ in \cref{app:Gammaxexpressions}. 

Figure \ref{fig:lorentzErrorRatesCombined}(a) compares the PLME instantaneous dephasing rates against those obtained from the numerically-exact evolution (see \cref{sec:QuasiExactRates}), for different values of the correlation time $\tau_c$. 
For the weak noise strength considered here ($\sigma / \Omega = 0.05$), there is an excellent agreement even for times that are large compared to $1 / \Omega$.

Similar to our analysis of quasistatic noise, we can study the approximation error (see \cref{eq:errorDnorm}) of the $2^{\rm nd}$-order PLME, and compare it to the simpler $0^{\rm th}$-order approximate master equation. This latter approximation equation is analogous to \cref{eq:quasi0th}: one first derives a dephasing dissipator ignoring the Rabi driving in the Hamiltonian, and only adds back the Rabi drive as a Hamiltonian term in the final master equation (see \cref{fig:system}(b)). Explicitly, this equation in the interaction frame reads:
\begin{align}
    \partial_{t} \rhoop(t) =&  2 \sigma^{2}\tauc \left( 1 - e^{-t/\tauc} \right) \nonumber \\
    & \times \mD{\cos (\Omega t ) \sz + \sin (\Omega t) \sy}{\rhoop}(t).
    \label{eq:lorentzZerothOrder}
\end{align}
Figure \ref{fig:lorentzErrorRatesCombined}(b) compares the approximation errors for these two approaches as a function of time, for three different choices of the noise correlation time $\tau_c$. We see that the PLME has significantly smaller approximation error than the naive $0^{\rm th}$-order equation, even for relatively short correlation times.

\subsection{Limiting long-time behavior}

For evolution times $t$ much longer than the noise correlation time, our instantaneous effective dephasing rates become time-independent, with the form:
\begin{align}
    \Gammapm(t) &\approx \sigma^{2} \tauc \left( \frac{1}{(\Omega \tauc)^{2}+1} \pm \frac{1}{\sqrt{(\Omega \tauc)^{2}+1}}\right).
    \label{eq:lorentzGammapmLongTime}
\end{align}
Note that in this limit, the second negative rate $\Gamma_{-}(t)$ remains non-zero.  

We can get further insight by considering the case of a weak drive, i.e., $\Omega \tauc \ll 1$, in addition to taking the long time limit.  Dropping terms of order $\bigoh{(\Omega \tauc)^{5}}$, we have 
\begin{align} 
\Gamma_{+}(t) &\approx  2\sigma^{2} \tauc \left(1-\frac{3 }{4} (\Omega \tauc)^{2} +\frac{11 }{16} (\Omega \tauc)^{4} \right),
\end{align}
\begin{align} 
\Gamma_{-}(t) &\approx - 2\sigma^{2}\tauc \left(\frac{1}{4} (\Omega \tauc)^{2}-\frac{5 }{16} (\Omega \tauc)^{4}\right),
\end{align}
and
\begin{align}
\Hren(t) &\approx  \sigma^{2}\tauc \left( \Omega \tauc- (\Omega \tauc)^{3} \right) \sx.
\end{align}
In this weak Rabi-drive limit, the negative rate
$\Gammam(t)$ vanishes as $(\Omega \tau_c)^2$, whereas the positive rate tends to the expected dephasing rate for simple undriven dephasing 
(i.e., it is determined by the noise spectral density at zero frequency).

In the opposite strong driving limit, where $\Omega \tauc \gg 1$, at leading order, both $\Gammapm(t)$ are needed to capture the qubit's dynamics properly. 
In particular, we see that in this limit (and at long times $t \gg \tau_c$):
\begin{align}
    \Gammapm(t) & \approx \pm \frac{\sigma^{2}}{\Omega} \left( 1 \pm \frac{1}{\Omega \tauc} - \frac{1}{2(\Omega\tauc)^{2}}  \right),
\end{align}
while 
\begin{align}
\Hren(t) &=\frac{ \sigma^{2}}{\Omega} \left(1 - \frac{1}{(\Omega \tauc)^{2}} + \frac{1}{(\Omega \tauc)^{4}} \right) \sx. 
\end{align}
In this case, $|\Gammap(t)| \approx |\Gammam(t)|$ and the dephasing jump operator associated with $\Gammam(t)$, can partially ``undo'' the loss generated by the jump operator of $\Gammap(t)$. Heuristically, this is also a manifestation of the same kind of continuous dynamical decoupling we discussed in our analysis of quasistatic noise:  the Rabi frequency is large enough to efficiently average away the comparatively slow dephasing noise.  


\section{Rabi driven qubit subject to $1/f$ noise}
\label{sec:1overfNoise}

We now investigate the PLME for a Rabi-driven qubit subject to classical noise with a $1/f$ spectral density, a kind of noise ubiquitous in many systems (e.g.,~superconducting qubits \cite{Ithier05}), with substantial ongoing efforts to mitigate its effects (see e.g., \cite{Didier2019acFluxSweet,huang2021engineering,huang2022high}). 
The noise spectral density of interest has the form
\begin{align}
    S[\omega]= 
\begin{cases}
    \sigma^{2} \frac{2\pi}{\abs{\omega}}, & \text{if } \omega_{l} \le \omega \le \omega_{h}\\
    0,              & \text{otherwise}
\end{cases}
    \label{eq:S1overfFreq}
\end{align}
with $\omega_{l}$ and $\omega_{h}$ representing the low and high frequency cutoffs respectively. In the time domain, the corresponding autocorrelation function reads
\begin{align}
    S(t) =& 2 \sigma^{2} \left(\Ci\left(\omega_{h} {| t |}\right)-\Ci\left(\omega_{l} {| t |}\right)\right) \\\nonumber 
    &\approx -2 \sigma^{2} \Ci\left(\omega_{l} {| t |}\right)
    \label{eq:1overfS}
\end{align}
with $\Ci(t) = - \int_{t}^{\infty} dt' \cos(t')/t'$. In the second line we have taken the limit of a large high frequency cutoff (i.e., $\omega_{h}t \rightarrow \infty$), something we will do throughout this section.  

\begin{figure}[t]
    \begin{subfigure}{1.0\columnwidth}
        \caption{}
        \includegraphics[width=1.0\columnwidth]{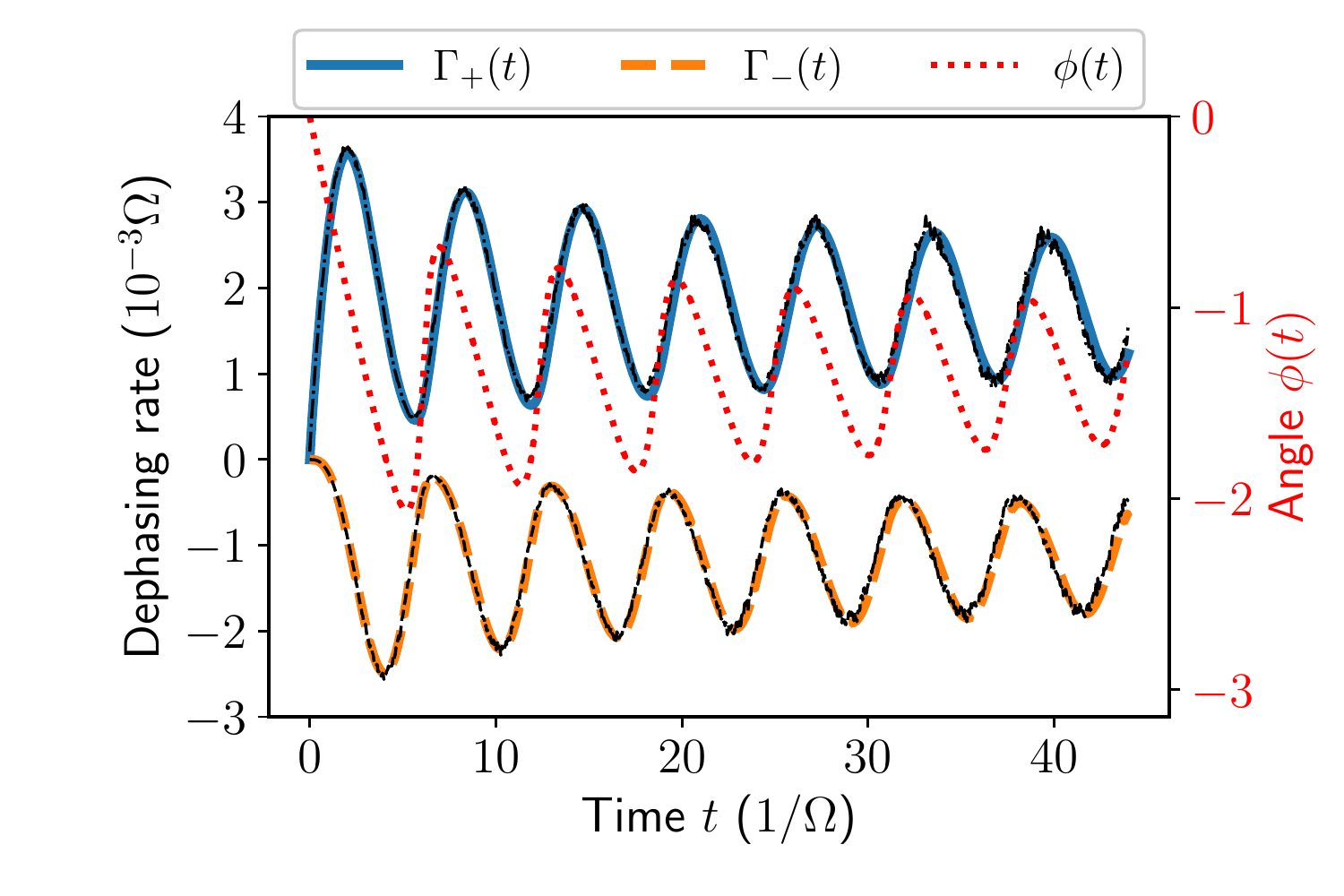}
    \end{subfigure}
    \begin{subfigure}{1.0\columnwidth}
        \caption{}
        \includegraphics[width=1.0\columnwidth]{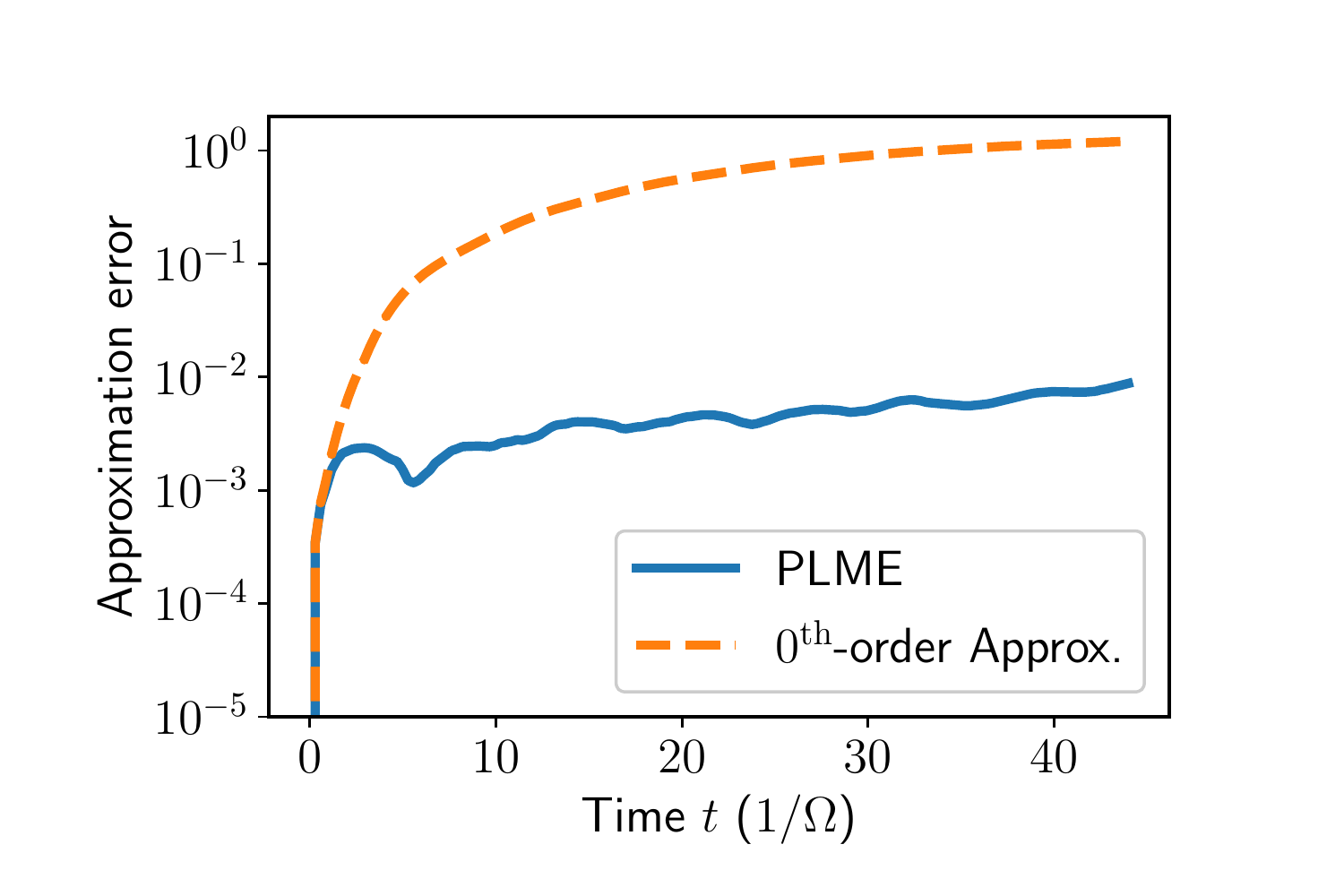} 
    \end{subfigure}
    \caption{
        Accuracy and performance of the $2^{\rm nd}$-order PLME for the case of $1/f$ noise.
        (a) PLME rates $\Gammapm(t)$ (blue and orange) along with rates obtained from the true qubit evolution (black), here estimated by averaging over 20,000 noise realizations. The red dotted curve (range given on the right axis) shows the oscillatory nature of the angle $\phi(t)$.
        (b) Approximation error [\cref{eq:errorDnorm}] of the $2^{\rm nd}$-order PLME (blue) and the $0^{\rm th}$-order approximation (orange) [\cref{eq:1overfZerothOrderSimple}]. 
        We stress that when numerically solving the PLME, we do not make the assumption that $\Omega \pm \omega_{l} \approx \Omega$, made to simplify the expressions in main text. 
    Parameters are $\sigma=0.01\Omega$ and $\omega_{l}=10^{-3}\Omega$.
    }
    \label{fig:1overfErrorRates}
\end{figure}

Once again, using the discussion in \cref{sec:cumulantGeneral,sec:drivenQubit}, we can calculate the rates $\Gammapm(t)$ as well as the noise-induced Hamiltonian term $\Hren(t)$ that enter the PLME.
We will also consider the case $\Omega\pm \omega_{l} \approx \Omega$ to arrive at simpler analytic expressions (though this simplification is not a requirement of our treatment). 
We find:
\begin{align} 
    \Gammapm(t) &=  \frac{2 \sigma^{2}}{\Omega} \left(-G_{u}(t) \mp  \textrm{sign}(G(t))  \sqrt{G_{u}(t)^{2}+G_{v}(t)^{2}}\right), 
\end{align}
and
\begin{align}
    \Hren &= \frac{2 \sigma^{2} }{\Omega}  G_{v}(t)  \sx,
    \label{eq:1overfHren}
\end{align}
where, for brevity, we have introduced
\begin{align} 
    G(t) &= \Ci\left(\omega_{l} t \right) \omega_{l} t -\sin \left(\omega_{l} t \right) ,\\
    G_{u}(t) &= \sin \left(\Omega t \right) \Ci\left(\omega_{l} t \right)-\Si\left(\Omega t \right), \\
    G_{v}(t) &= \cos \left(\Omega t \right) \Ci\left(\omega_{l} t \right)+\ln \left(\frac{\Omega}{\omega_{l}}\right)-\Ci\left(\Omega t \right),
\end{align}
with $\Si(t) = \int_{0}^{t} dt' \sin(t')/t' $. 
The angle $\phi(t)$ satisfies 
\begin{align}
    \cos \phi(t) &= \frac{G_{u}}{\sqrt{G_{u}^{2} + G_{v}^{2}}}, 
    \quad
    \sin \phi(t) = \frac{G_{v}}{\sqrt{G_{u}^{2} + G_{v}^{2}}}.
    \label{eq:1overfPhi}
\end{align}

In \cref{fig:1overfErrorRates}(a), besides plotting an explicit form of $\phi(t)$, we also show that the analytical approximations for the rates $\Gammapm(t)$ exhibit good agreement with rates obtained from the numerically-exact evolution.  As discussed in \cref{sec:PLMEFourthOrder}, at 
$4^{\rm th}$-order in the cumulant expansion one also obtain a time-dependent $x$-dephasing rate $\Gamma_{x}(t)$; its expression is unwieldy so we do not include it here. For the parameters used in \cref{fig:1overfErrorRates}, $\Gamma_{x}(t)$ is much smaller than the rates $\Gammapm(t)$ (hence we do not show it on this plot).

Figure \ref{fig:1overfErrorRates}(b) exhibits a performance comparison of the $2^{\rm nd}$-order PLME and a $0^{\rm th}$-order approximation master equation, 
which (in the interaction frame) reads
\begin{align}
    \partial \rhoop(t) =& \Gamma_{0^{\rm th}}(t) \mD{\cos (\Omega t ) \sz + \sin (\Omega t) \sy}{\rhoop}(t).
    \label{eq:1overfZerothOrderSimple}
\end{align}
with 
\begin{align}
     \Gamma_{0^{\rm th}}(t) =&   
     4 \sigma^{2}t \left( \frac{\sin(\omega_{l} t)}{\omega_{l} t}  - \Ci\left(t \omega_{l}\right) \right) \nonumber \\
     \approx & 4 \sigma^{2} t \ln(\omega_{l}t),
    \label{eq:1overfRate}
\end{align}
where in the last line, for completeness, we show the commonly used approximation obtained by only keeping the leading-order contribution in $\omega_{l}t$ (note, that in simulations we use the form shown in the first line of \cref{eq:1overfRate}).

From the plot, it is clear that the PLME is a more accurate description of the dynamics than the $0^{\rm th}$-order approximation. 
As was the case with the other highly-correlated noise types, the $0^{\rm th}$-order approximation, completely misses the coherent renormalization Hamiltonian $\Hren(t)$, the jump operator associated with the negative rate $\Gammam(t)$, as well as, overestimates the decoherence resulting from the positive rate $\Gammap(t)$.

It is tempting to assume that the effects of $1/f$ noise are similar to that of quasistatic noise, as both are correlated over long time-scales. Further, in the free-induction decay limit (where $\Omega \rightarrow 0$), they both give rise to dephasing that depends exponentially on $t^2$.  Our analysis shows that the correspondence between these two kinds of noise is far from exact when we include a Rabi drive. For quasistatic noise (at the level of $2^{\rm nd}$-order cumulant expansion contributions), the two dephasing rates $\Gammapm(t)$ are non-decaying oscillatory functions, and have equal maximum amplitudes.  Further, the angle $\phi(t)$ simply advances linearly with time.

Our results for $1/f$ noise show that the symmetry in amplitudes of the rates $\Gammapm(t)$ is broken. In particular, from \cref{fig:1overfErrorRates}(a), we see that in contrast to quasistatic noise, for $1/f$ noise
the instantaneous dephasing rate $\Gamma_{+}(t)$ is more dominant (in amplitude) than the negative rate $\Gamma_{-}(t)$.  Further, the oscillations in both $\Gammapm(t)$ as well as the angle $\phi(t)$ get smaller as time evolves (on relevant times scales, where $\omega_{l}t \ll 1$). 
Accounting for these types of differences correctly may prove to be important in applications where accurate modeling of the effects of these different environment types on the qubit dynamics is required.

\section{Conclusions}
\label{sec:Conclusions}

This work introduces and derives a new time-local pseudo-Lindblad master equation (PLME) that accurately models the effects of Gaussian non-Markovian classical noise on driven quantum systems.  The approach is based on a generalized cumulant expansion, and yields a master equation that can be written in terms of time-dependent Hamiltonian and Lindblad dissipators.  We find that the combination of non-Markovianity and driving leads generically to instantaneous dephasing rates that can be negative.  Through a detailed comparison against numerics for a Rabi-driven qubit, we find that the PLME can give an extremely accurate description of dynamics over timescales typically required to perform qubit control operations.  We considered three forms of non-Markovian noise:  quasistatic, Lorentzian as well as $1/f$, which is often dominant in various physical scenarios.  As our PLME is time-local, and requires no stochastic averaging, it 
may help simplify and speed up accurate modeling of noisy quantum systems, that may otherwise be notoriously difficult to simulate.

\begin{acknowledgments}
    This work was primary supported by the Department of Energy BES 
    Quantum Information Science Program under Award No.~DE-SC002015.  
    A.S. is supported by a Chicago Prize Postdoctoral Fellowship in Theoretical Quantum Science.
    J.K. was supported by the ARO under Grant No.~W911NF-1910016.  
    A. A. C. acknowledges support from the Simons Foundation through a Simons Investigator award (Grant No. 669487, A. A. C.).
    
\end{acknowledgments}


\begin{appendix}


\section{Generalized cumulant expansion}
\label{app:CumulantBasics}

In this Appendix, we briefly outline the derivation of the cumulant expansion that we use to obtain the PLME.  While many different versions and derivations of such expansions exist, we follow a particularly physically transparent one that is similar to Budimir and Skinner \cite{Budimir_Skinner_1987}.  

Our starting point is the general stochastic master equation shown in \cref{eq:htotal}. We can formally solve it (for any given noise realization) as:
\begin{align}  
    \rhoop_{\eta}(t) & = 
        \Tsop  \exp \left(
             \int_0^t dt' \eta(t') \mathcal{L}(t')
         \right) \rhoop_{\eta}(0)
\end{align}
where $\Tsop$ represents the time-ordering operator, and 
where we have defined the superoperator $\mathcal{L}(t)$ via
\begin{align}
    \mathcal{L}(t) \hat{Q} = -i \left[ g(t) \hat{A}(t) , \hat{Q}\right],
    \label{eq:LsuperDef}
\end{align}
for some operator $\Qop$.
Next, one can formally perform an exact average over the Gaussian noise $\eta(t')$ yielding
\begin{align}
    \rhoop(t) \equiv \mathcal{U}(t) \rhoop(0).  
\end{align}
with
\begin{align}
    \mathcal{U}(t) & = \Tsop  \exp \left(
    \frac{1}{2} \int_0^t \int_0^t dt_1 dt_2 
        S(t_1 - t_2) 
        \mathcal{L}(t_1) \mathcal{L}(t_2) 
    \right),
\end{align}
where $S(t)$ represents the spectral density of the bath. 
We next assume that $\mathcal{U}(t)$ is invertible, and use the above equation to derive a formally exact time-local equation for $\partial_t \rhoop(t)$
\begin{align}
    \partial_t \rhoop(t) = \left[ \partial_t \mathcal{U}(t) \right] \mathcal{U}^{-1}(t) \rhoop(t) 
    \equiv \Rsop(t) \rhoop(t)
\end{align}

The generalized cumulant expansion now arises from a perturbative expansion of the superoperator 
$\Rsop(t)$ in powers of the noise, i.e., in powers of $\eta(t')\mathcal{L}(t')$, thus letting us write
\begin{align}
    \Rsop(t) &= \RsopTwo(t) + \RsopFour(t) + \dots 
    \label{eq:Rexp}
\end{align}
The leading non-zero term is at $2^{\rm nd}$-order in the noise. 
At this order, $\mathcal{U}^{-1}$ is approximated as unity, and one obtains \cref{eq:meGenR2} from the main text.  As discussed, this is exactly the form of a Redfield master equation.  This equation is also exact in the special cases where either $\hat{A}(t)$ commutes with itself at different times, or when the noise is delta correlated.  In these cases, there is a cancellation at all higher orders between the expansion of $\mathcal{U}(t)$ and that of $\mathcal{U}^{-1}(t)$.

For the general case, the next non-vanishing contribution comes at $4^{\rm th}$-order in the noise, i.e.,~the fourth cumulant. It takes the general form:
\begin{align}
    \RsopFour(t) =& \Lsop(t)  \int_{0}^{t} dt_{1} \int_{0}^{t_{1}} dt_{2} \int_{0}^{t_{2}} dt_{3} \\\nonumber
&\times \bigg( S_{02} S_{13} \comm{\Lsop(t_{1})}{\Lsop(t_{2})} \Lsop(t_{3}) \\\nonumber 
& + S_{03} S_{12} \Lsop(t_{1}) \comm{\Lsop(t_{2})}{\Lsop(t_{3})}  \\\nonumber 
& + S_{03} S_{12} \comm{\Lsop(t_{1})}{\Lsop(t_{3})} \Lsop(t_{2}) \bigg),
\end{align}
where, for notational convenience, we have introduced $S_{ij}$ to mean $S(t_{i} - t_{j})$, with $t_{0}=t$. 
It is easy to see that using \cref{eq:LsuperDef}, the above superoperator commutators, can be readily written in terms of commutators of the system Hamiltonian, and hence of $\Aop(t)$. 
The non-commutativity of $\hat{A}(t)$ at different times means that $\mathcal{U}^{-1}(t)$ does not cancel all contributions at this order; we see explicitly that all of the terms above involve commutators of the superoperator $\mathcal{L}(t')$.

\section{Expectation value evolution}
\label{app:expvals}

\begin{figure}[t]
    \begin{subfigure}{1.0\columnwidth}
        \caption{}
        \includegraphics[width=1.0\columnwidth]{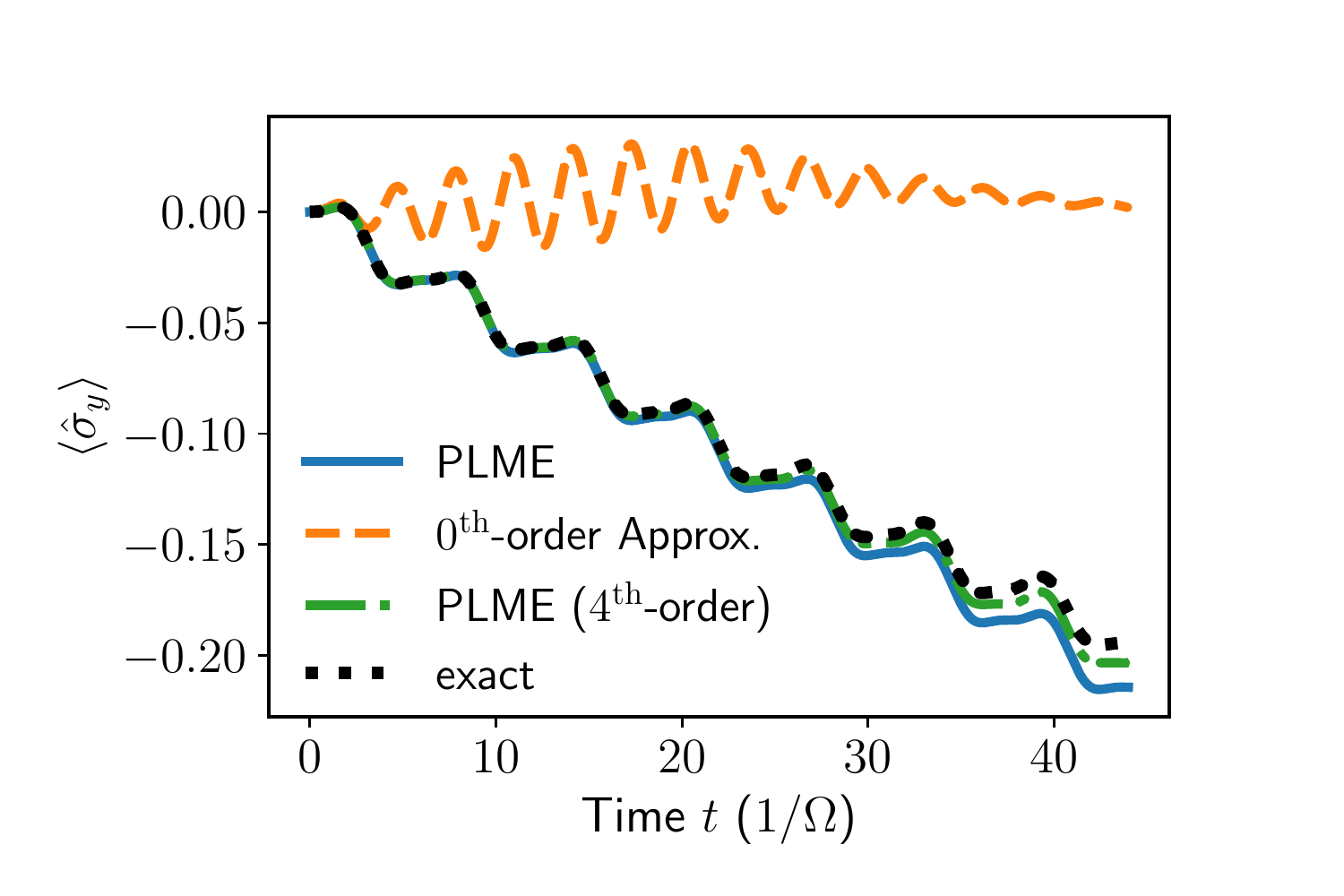}
    \end{subfigure}
    \begin{subfigure}{1.0\columnwidth}
        \caption{}
    \includegraphics[width=1.0\columnwidth]{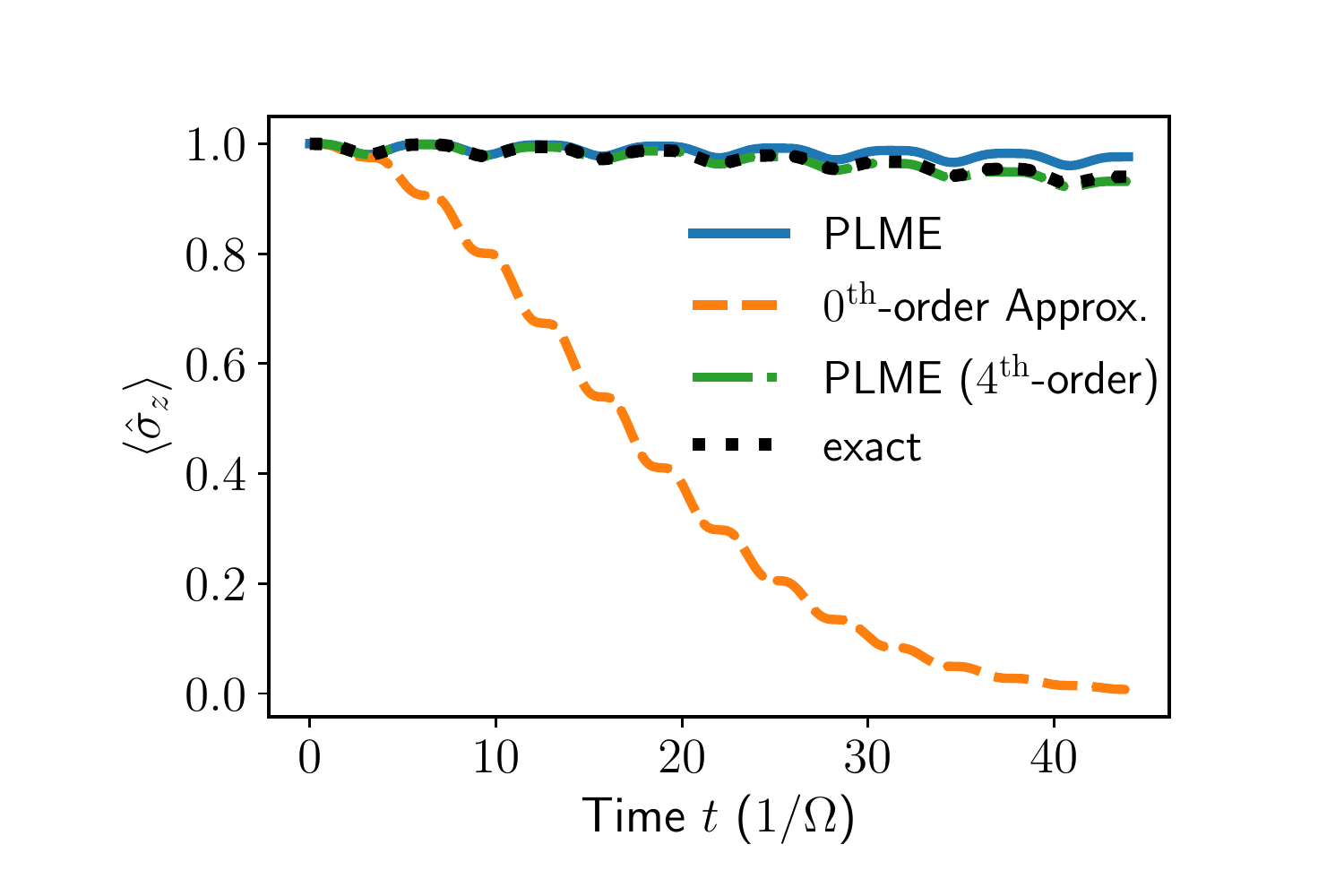}
    \end{subfigure}
    \caption{ 
        Evolution of the qubit affected by quasistatic noise, shown in terms of expectation values: panel (a) $\ave{\sy}$ and (b) $\ave{\sz}$, as a function of time. 
       Different curves are obtained using the PLME when $2^{\rm nd}$-order (blue, solid) or $4^{\rm th}$-order (green, dash-dotted) terms are kept in the cumulant expansion, the $0^{\rm th}$-order approximation (orange, dashed), as well as numerically-exact evolution (black, dotted). The qubit starts in an initial state $\ket{0}$ (positive energy eigenstate of $\sz$), and the simulation is performed in the interaction frame (see~\cref{eq:AopDrivenQubit}), with $\sigma/\Omega=0.05$. 
    }
    \label{fig:expValsQuasistatic}
\end{figure}

\begin{figure*}
\captionsetup[subfigure]{slc=off,margin={10pt,10pt}}
    \begin{subfigure}{0.38\textwidth}
        \caption{}
        \includegraphics[width=1.0\textwidth]{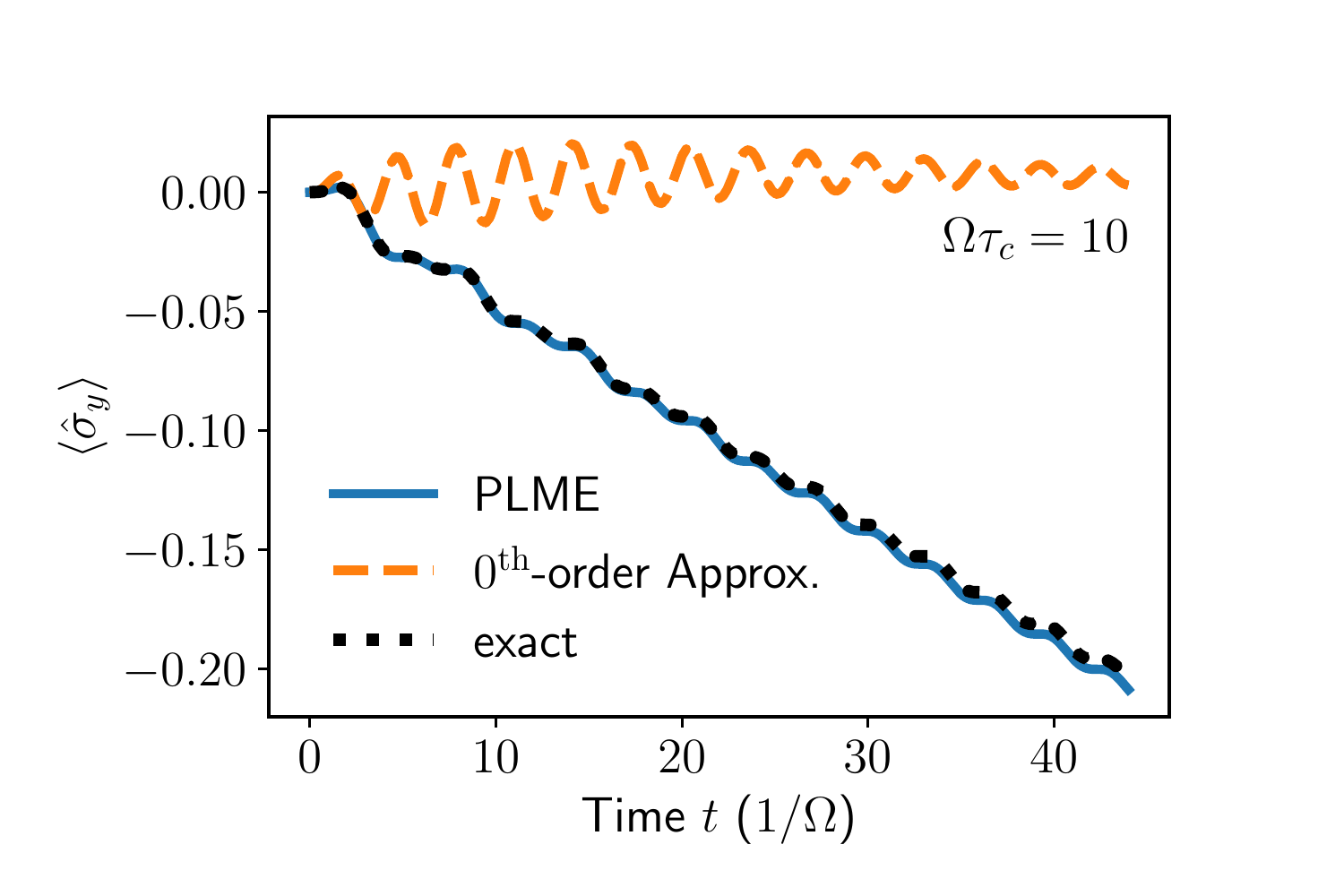}
    \end{subfigure}
    \hspace{-1.2cm}
    \begin{subfigure}{0.38\textwidth}
        \caption{}
        \includegraphics[width=1.0\textwidth]{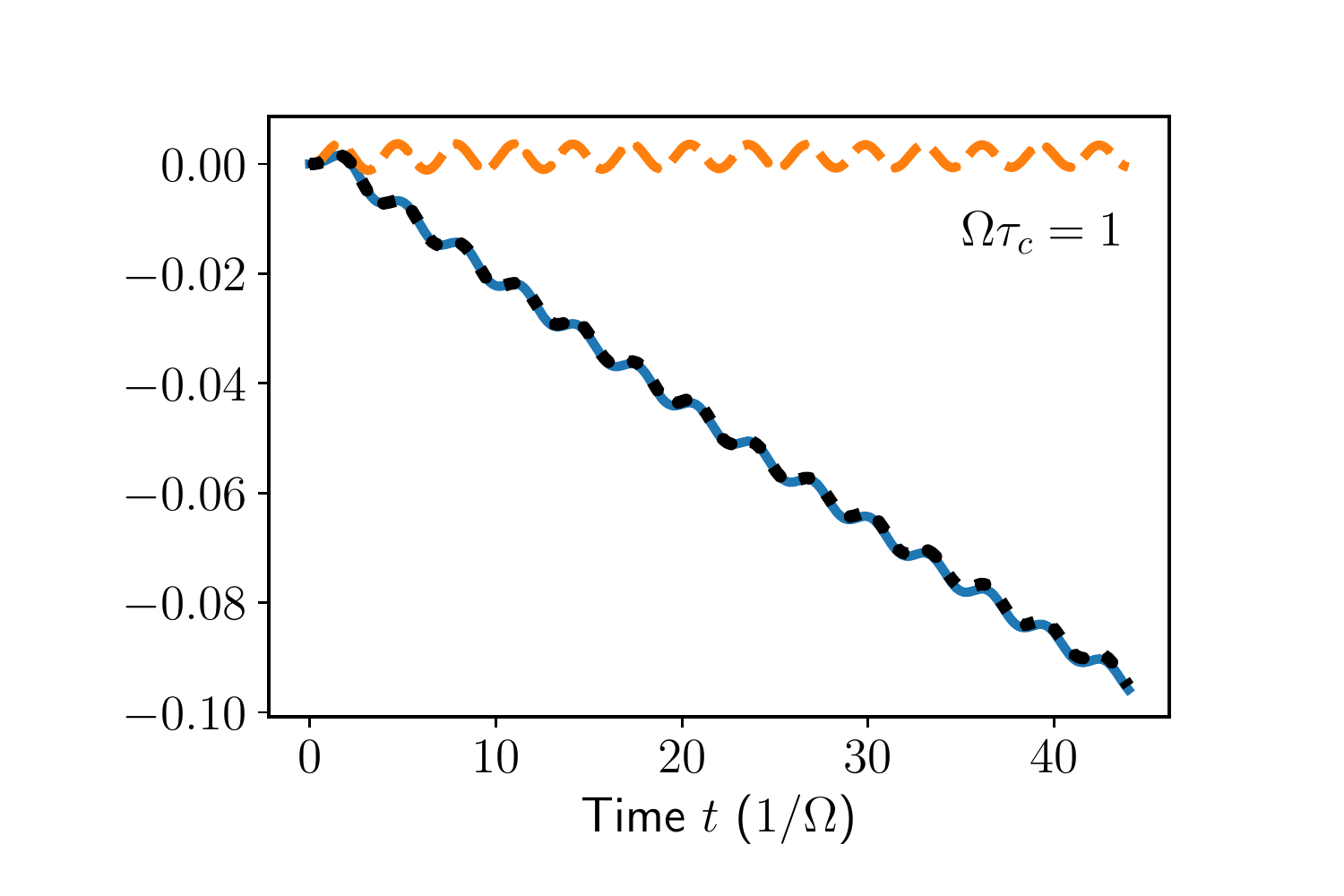}
    \end{subfigure}
    \hspace{-1.2cm}
    \begin{subfigure}{0.38\textwidth}
        \caption{}
        \includegraphics[width=1.0\textwidth]{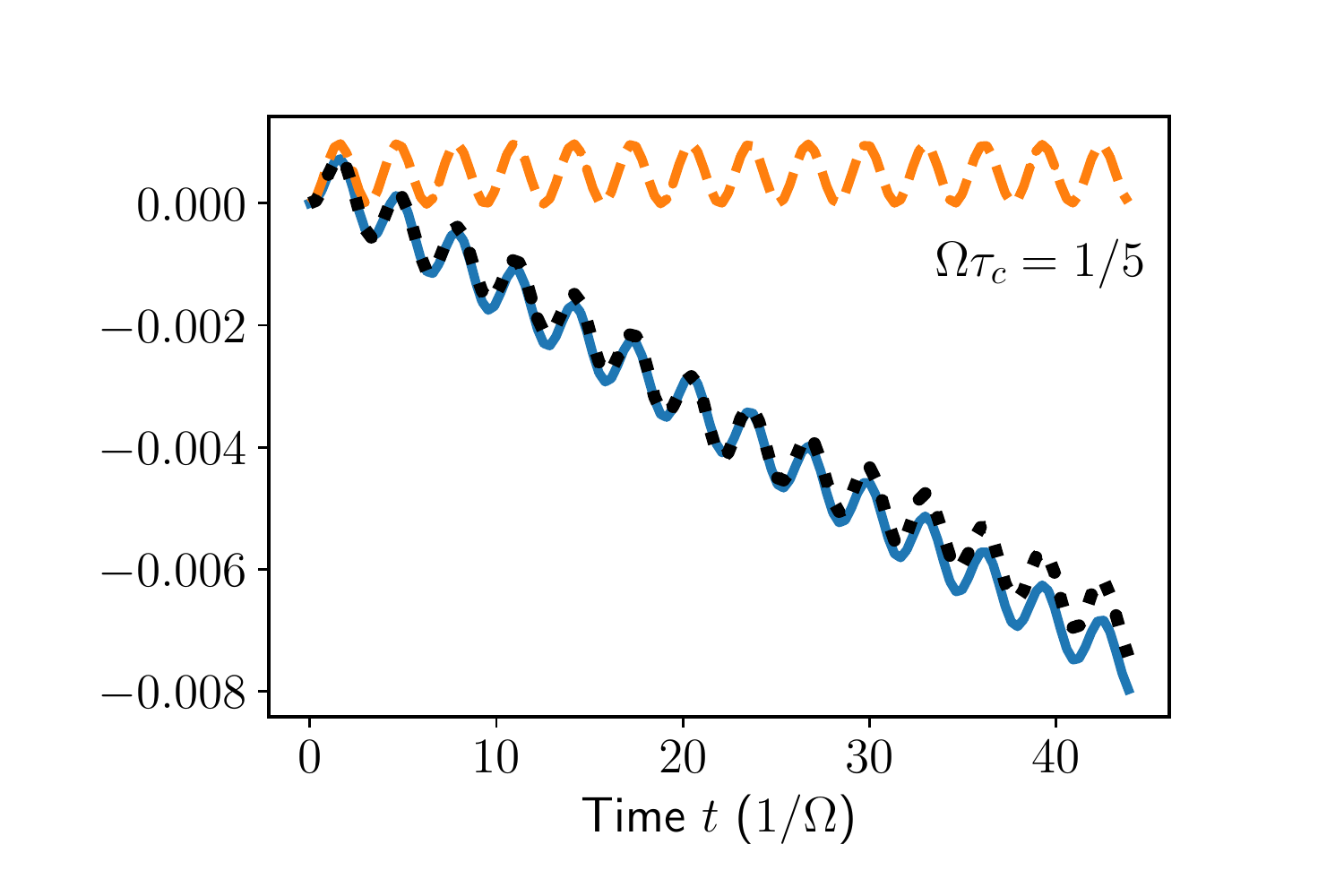}
    \end{subfigure}
    \\
    \begin{subfigure}{0.38\textwidth}
        \caption{}
        \includegraphics[width=1.0\textwidth]{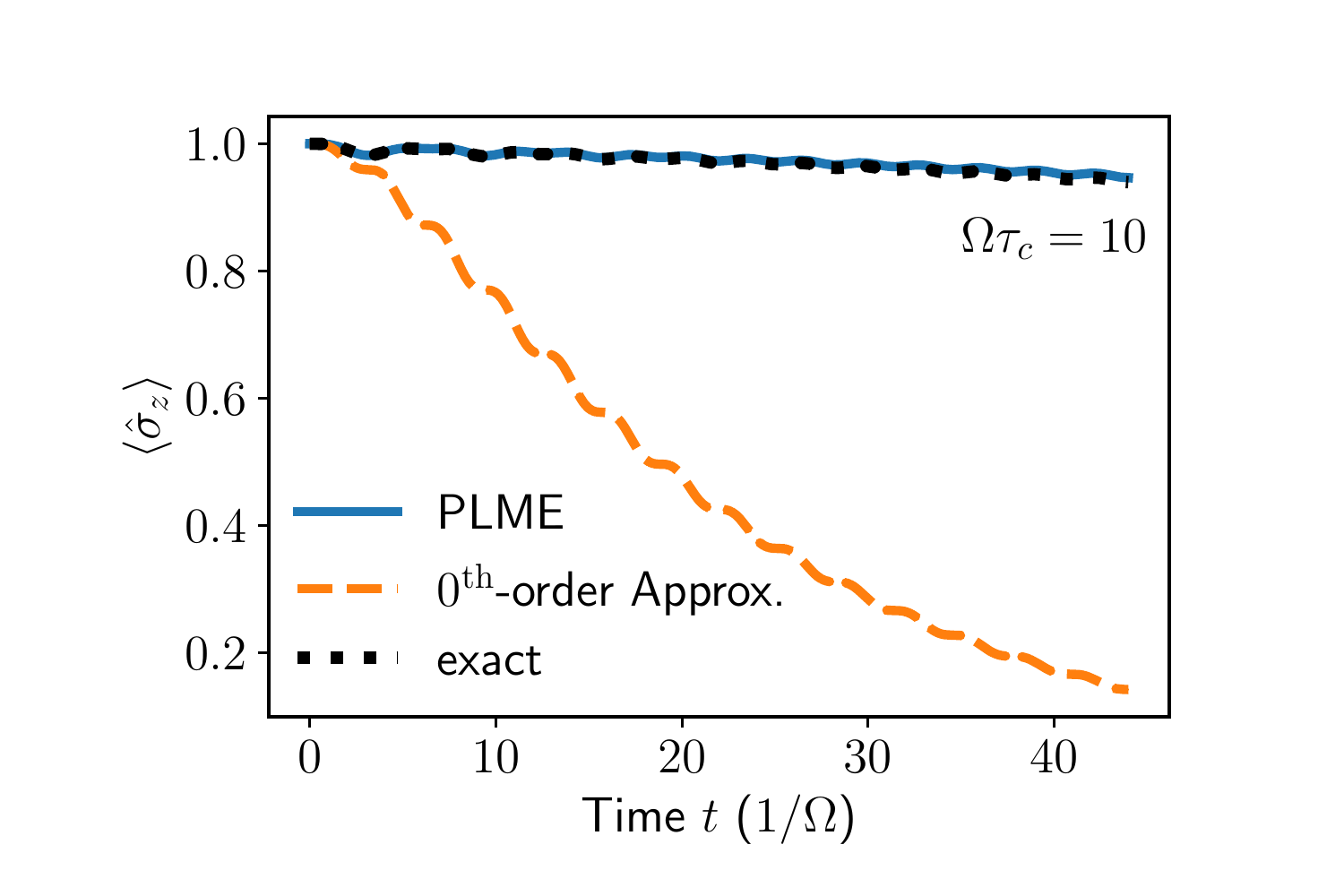}
    \end{subfigure}
    \hspace{-1.2cm}
    \begin{subfigure}{0.38\textwidth}
        \caption{}
        \includegraphics[width=1.0\textwidth]{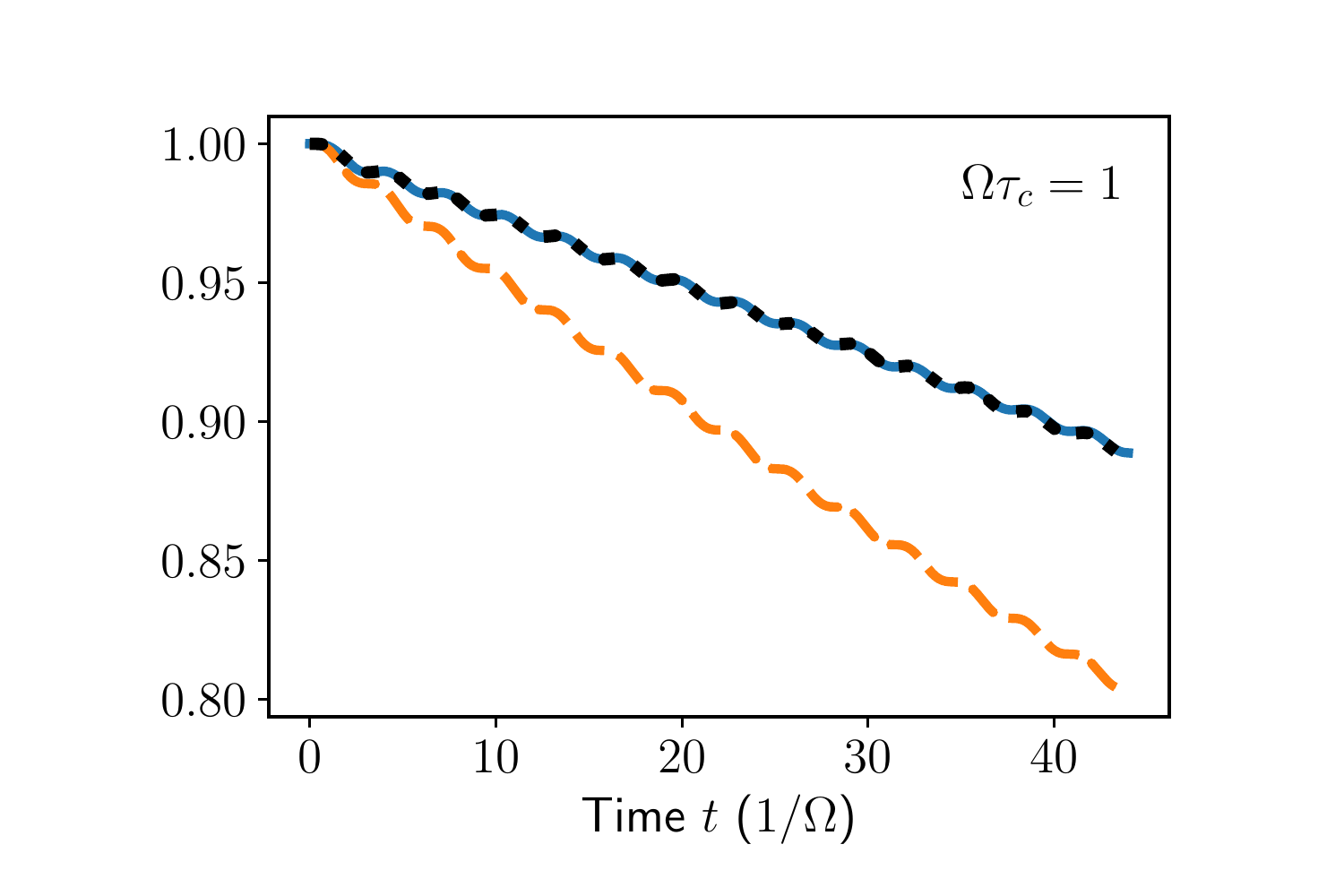}
    \end{subfigure}
    \hspace{-1.2cm}
    \begin{subfigure}{0.38\textwidth}
        \caption{}
        \includegraphics[width=1.0\textwidth]{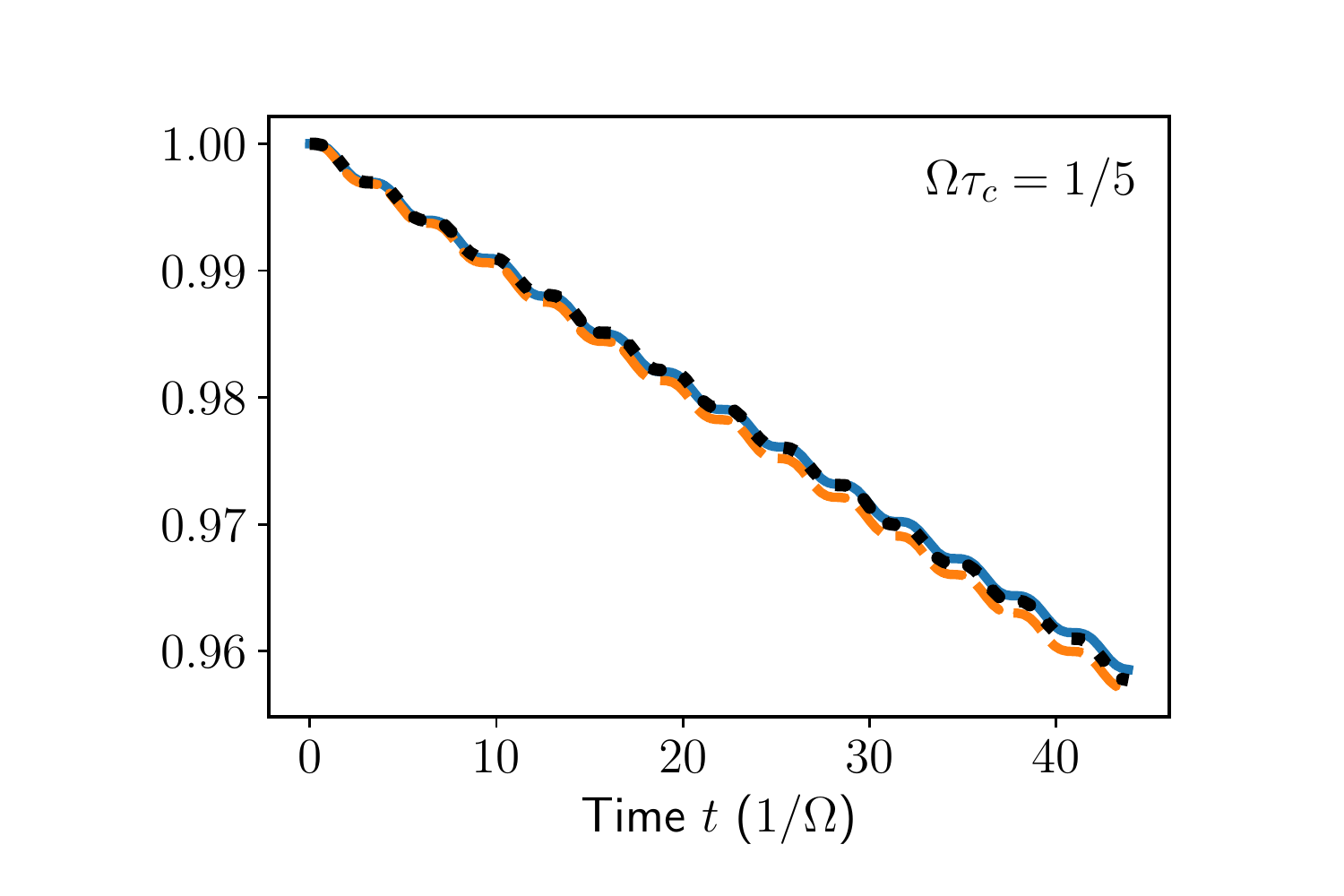}
    \end{subfigure}
    \caption{ 
        Evolution of the qubit affected by Lorentzian noise, shown in terms of expectation values: panels (a-c) show $\ave{\sy}$, while (d-f) $\ave{\sz}$, as a function of time. 
        Different curves are obtained using the PLME (blue, solid), the $0^{\rm th}$-order approximation (orange, dashed), as well as numerically-exact evolution (black, dotted), obtained from averaging over 20000 noise realizations. 
        The correlation time $\tauc$ is largest in the left panels and decreases towards the right. In particular, we have in (a), (d): $\Omega \tauc = 10$, (b), (e): $\Omega \tauc =1$ and (c), (f): $\Omega \tauc=1/5$. The qubit starts in an initial state $\ket{0}$ (positive energy eigenstate of $\sz$), and the simulation is performed in the interaction frame (see~\cref{eq:AopDrivenQubit}), with $\sigma=0.05\Omega$. 
    }
    \label{fig:expValsLorentzian}
\end{figure*}

In the main text, in order to compare the accuracy of the PLME and the $0^{\rm th}$-order approximation to the \textit{numerically-exact} evolution, we studied the approximation error $\epsilon(t)$ (see \cref{eq:errorDnorm}). Such a measure is conservative, but gives a useful lower bound to how well a given modeling approach performs.

\begin{figure}[t]
    \begin{subfigure}{1.0\columnwidth}
        \caption{}
        \includegraphics[width=1.0\columnwidth]{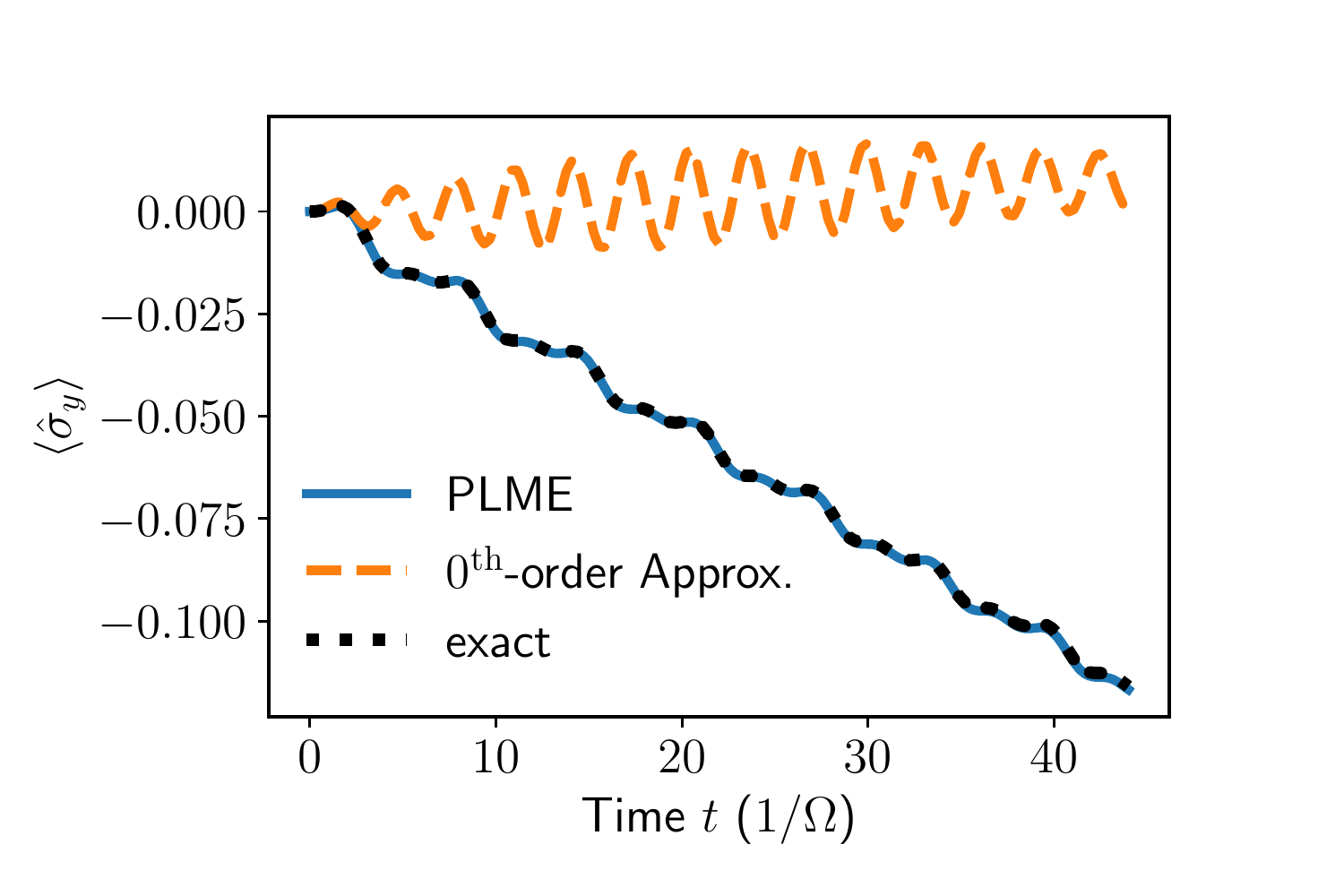}
    \end{subfigure}
    \begin{subfigure}{1.0\columnwidth}
        \caption{}
        \includegraphics[width=1.0\columnwidth]{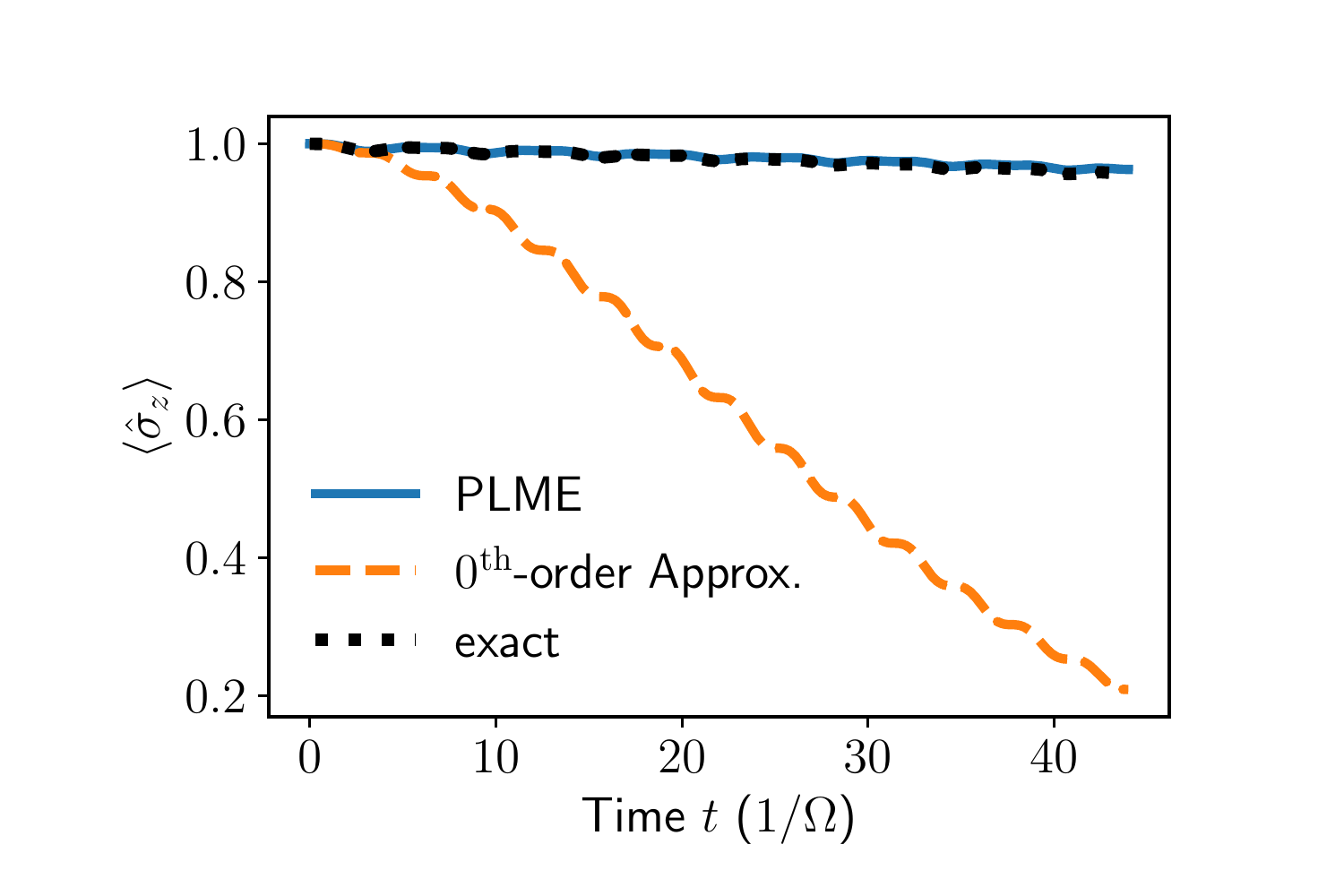}
    \end{subfigure}
    \caption{
        Evolution of the qubit affected by $1/f$ noise, shown in terms of expectation values: panel (a) $\ave{\sy}$ and (b) $\ave{\sz}$, as a function of time. 
        Different curves are obtained using the PLME (blue, solid), the $0^{\rm th}$-order approximation (orange, dashed), as well as numerically-exact evolution (black, dotted), obtained from averaging over 20000 noise realizations. The qubit starts in an initial state $\ket{0}$ (positive energy eigenstate of $\sz$), and the simulation is performed in the interaction frame (see~\cref{eq:AopDrivenQubit}), with $\sigma=0.01\Omega$ and $\omega_{l}=10^{-3}\Omega$.
    }
    \label{fig:expVals1overf}
\end{figure}

To present the reader with a little more insight, in this Appendix we also show the expectation values $\ave{\sy}$ and $\ave{\sz}$ (noting that $\ave{\sx}$ has more trivial dynamics and thus for brevity, is not presented explicitly) as a function of time, of a particular initial state, here taken as $\ket{0}$ (positive energy eigenstate of $\sz$). Figures \ref{fig:expValsQuasistatic}, \ref{fig:expValsLorentzian} and $\ref{fig:expVals1overf}$ show the results for quasistatic, Lorentizan as well as $1/f$ noise, respectively. In each case, the evolution is obtained in the interaction frame (see discussed around \cref{eq:AopDrivenQubit}), and thus in the limit of vanishing noise strength (i.e., $\sigma \rightarrow 0$) one should expect both $\ave{\sy}$ and $\ave{\sz}$ to stay constant as time evolves. We once again highlight the fact that the very model of a driven qubit we consider here, can be conveniently interpreted as a case of a continues dynamical-decoupling protocol: the transverse drive perpetually flips the qubit, and with each half-oscillation the effects of longitudinal noise can be partly ``mitigated'' (to what extent that happens largely depends on the noise type). 

The shown plots complement our main text discussion: namely, when the noise is highly non-Markovian, the $0^{\rm th}$-order approximation provides a poor description of the evolution, while the PLME, in contrast, closely captures it both qualitatively and quantitatively. This is directly observed for the quasistatic, $1/f$, as well as long-correlated (i.e., $\Omega \tau_{c}=10$) Lorentzian noise. As the noise correlation time decreases (see \cref{fig:expValsLorentzian} panels (b), (e), and (c), (f)), the $0^{\rm th}$-order approximation starts to perform better, i.e., as $\tau_{c}\rightarrow 0$, the system tends towards the white-noise limit, one where both approximations (PLME and the $0^{\rm th}$-order) reach parity (see discussion in \cref{sec:cumulantGeneral} and \cref{sec:LorentzianNoise}). Finally we point out that, naturally, the exact form of the curves in \cref{fig:expValsQuasistatic,fig:expValsLorentzian,fig:expVals1overf} will change as one chooses a different initial state, the general trend discussed above, however, should hold even in those different cases.

\section{Detailed study of PLME positivity for quasistatic noise}
\label{app:Positivity}

In this Appendix we address the question of whether the evolution that our PLME leads to, is actually physical, i.e., can be described by a completely positive and trace-preserving (CPTP) map \cite{Nielsen00,Breuer2016RMP}? While in the case of a standard Lindblad master equation (even with time-dependent, but non-negative rates), this is guaranteed \cite{Gorini_1976,Breuer2016RMP}, we find that the PLME presented here, at times, can lead to dynamics that is not always completely positive. Even though this may seem as especially problematic, in this Appendix, we argue that this non-physicality of the $2^{\rm{nd}}$- and $4^{\rm{th}}$-order PLME evolution may be only relevant for some limited subset of initial states, and final evolution times, and hence in practice may not be especially problematic. 

\begin{figure}[t]
    \centering
    \includegraphics[width=0.46\textwidth]{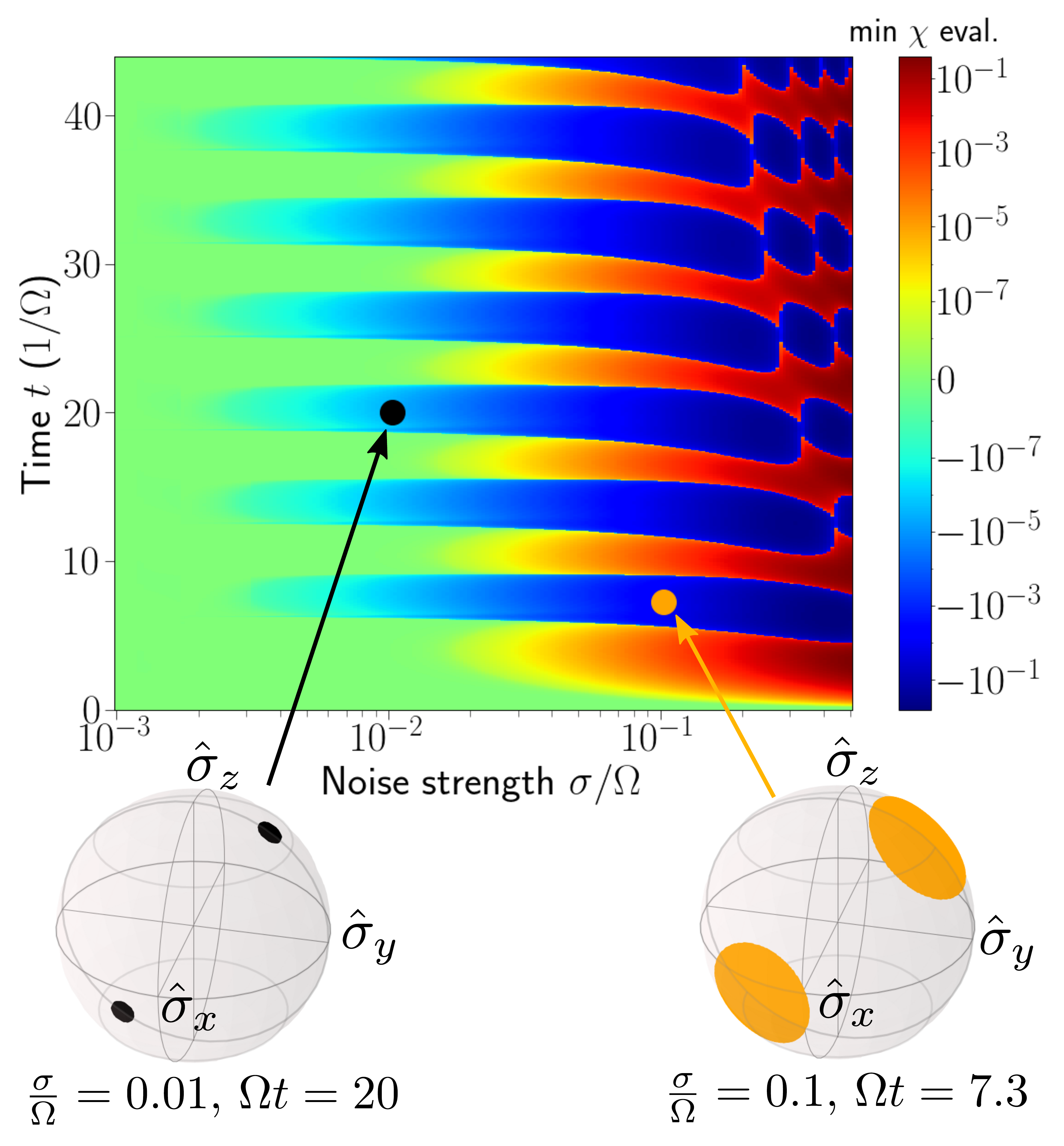}
     \caption{  
        The smallest eigenvalue of the process matrix $\chi(t)$ associated with the PLME evolution under quasistatic noise, plotted as a function of evolution time and the noise strength $\sigma$. Negative regions correspond to evolution that is non-physical. For two particular configurations (black and orange dots on the density plot), we show a subset of initial pure states (``patches'' on Bloch spheres) that result in non-physical final states. As one might expect, this subset of states is smaller at weaker noise strengths. 
    }
    \label{fig:quasiPositivity}
\end{figure}

To explore this aspect of our theory further, we consider a Rabi-drive qubit, subject to quasistatic noise (see \cref{sec:quasistatic} for a detailed discussion) and describe the map associated with the $2^{\rm th}$-order PLME evolution in terms of a (numerically obtained) process matrix (often referred to as the $\chi$-matrix) which reads
\begin{align}
    \mE_{t} (\rho) \rightarrow \sum_{n,m} \chi_{nm}(t) \Pop_{n} \rhoop \Pop_{m},
\end{align}
where $\Pop_{k} \in \{\iden, \sx, \sy, \sz \}$ form an operator-basis for a single qubit. 
Any map is guaranteed to be completely positive as long as the corresponding $\chi$-matrix is positive-semidefinite \cite{Breuer2016RMP}.
Hence, in \cref{fig:quasiPositivity} we show a plot of the smallest eigenvalue of $\chi$ as a function of the evolution time and noise strength. It is clear that there are certain times, at which the evolution from the PLME dynamics is not physical. As expected, at those times, the non-complete-positivity is more of a problem at higher noise strengths. This is because the leading-order error that our approach makes in modeling the \textit{true} evolution, scales as the quartic power of the noise-strength $\sigma$ - see discussion in \cref{sec:quasistatic}. We also observe a periodic behavior that follows the Rabi drive period $\sim 1 / \Omega$. In particular the PLME map starts to be non-completely-positive approximately at times $\sim 2\pi m/ \Omega$, with $m \in \mathbb{Z}^{+}$, and stays non-physical for roughly a half of a Rabi period.
This behavior is mainly due the interplay between the dissipators associated with rates $\Gammapm(t)$. In particular, for some set of initial states, the negative rate may be a more dominant of the two, dictating much of the evolution, which can in fact at times ``stretch'' the qubit's state vector to be larger than 1, resulting in a potentially non-physical final state. Naturally this should not happen (and does not happen in the true dynamics of the system), however, our analytic form of $\Gammam(t)$ is only an approximation to the \textit{true} rates of the instantaneous Liouvillian (see \cref{fig:quasiRates} of the main text).

In \cref{fig:quasiPositivity} we also choose two particular maps (marked by black an orange dots on the density plot) and show on corresponding Bloch spheres a set of pure initial states (colored patches) that would lead to non-physical final states (i.e., density matrices that are not positive-semidefinite) when the given map is applied. We see that the subset of these problematic states is typically expected to be small, as long as the noise strength is not too large. 

We stress that this type of artifact of our PLME is not unique when it comes to modeling the effects of highly non-Markovian noise. Other methods (for example, see \cite{Hartmann_Strunz_2020}), can also result in dynamics that leads to final states which may not be fully physical. Ultimately, as long as the error made relative to the true system evolution is small enough, an approximation obtained from our PLME may still be a preferred over the much more widely used approaches, that can result physical final states, but at a cost of much less accurate dynamics.
Furthermore, the evolution obtained with a PLME can be readily made physical, by projecting the final evolved state onto a ``closest'' physical state. Alternatively, a similar projection procedure can be also performed at the process map level. Such approaches can be readily implemented, and in practice can have little effect on the dominant error measures as long as the noise is not too strong.

\section{Dephasing rate $\Gammax(t)$ for Lorentzian noise}
\label{app:Gammaxexpressions}

In this brief Appendix, we write down the complete expression for the $\Gammax(t)$ $x$-dephasing rate (calculated by keeping $4^{\rm th}$-order terms in the cumulant expansion) for the case of Lorentzian noise. It reads
\begin{align} 
    \Gammax(t) =& 
    2 \sigma^{4} \bigg( 
    \frac{ \tau_{c}^{5} \Omega^{2} \left(\Omega^{2} \tau_{c}^{2}+5\right)}{\left(\Omega^{2} \tau_{c}^{2}+1\right)^{3}}
    -\frac{\tau_{c}^{3}   e^{-\frac{2 t}{\tau_{c}}}}{\left(\Omega^{2} \tau_{c}^{2}+1\right)^{3}} \beta_{1} \nonumber \\
    & -\frac{2   \tau_{c}^{2} e^{-\frac{t}{\tau_{c}}}}{\left(\Omega^{2} \tau_{c}^{2}+1\right)^{2} \Omega} \beta_{2} \bigg)
\end{align}
with
\begin{align} 
    \beta_{1} =& \Omega^{4} \tau_{c}^{4}-\sin \left(2 t \Omega \right) \Omega^{3} \tau_{c}^{3}+\Omega^{2} \left(3 \cos \left(2 t \Omega \right)+2\right) \tau_{c}^{2} \nonumber \\
    &+3 \sin \left(2 t \Omega \right) \Omega  \tau_{c} -\cos \left(2 t \Omega \right)+1,
\end{align}
and
\begin{align} 
\beta_{2} =& \Omega^{2}\tau_{c}^{2} \left(\cos \left(t \Omega \right) \Omega  t +\sin \left(t \Omega \right)\right) +2 \Omega^{2} t \tau_{c} \sin \left(t \Omega \right)  \nonumber \\
&-  \Omega  t  \cos \left(t \Omega \right) +\sin \left(t \Omega \right).
\end{align}
As we expect, it scales with the fourth power of the noise strength $\sigma$. 

\end{appendix}


\bibliographystyle{quantum}
\bibliography{library}


\end{document}